\documentclass[11pt,a4paper]{article}
\pdfoutput=1
\usepackage{jheppub} 
\usepackage{amsmath, amsfonts, amssymb}
\usepackage{graphicx, subfig, color, verbatim}
\usepackage[dvipsnames]{xcolor}
\usepackage{float}

\preprint{UUITP-51/21}

\title{ One-loop Gluon Amplitudes in AdS}
\author[a]{Luis F. Alday,}
\author[b]{Agnese Bissi,}
\author[c,d]{Xinan Zhou}
\affiliation[a]{Mathematical Institute, University of Oxford, Andrew Wiles Building, Radcliffe Observatory Quarter, Woodstock Road, Oxford, OX2 6GG, U.K.}
\affiliation[b]{Department of Physics and Astronomy, Uppsala University, Box 516, SE-751 20 Uppsala, Sweden.}
\affiliation[c]{Kavli Institute for Theoretical Sciences, University of Chinese Academy of Sciences, Beijing 100190, China.}
\affiliation[d]{Princeton Center for Theoretical Science, Princeton University, Princeton, NJ 08544, USA.}
\emailAdd{alday@maths.ox.ac.uk\\  \hskip 42pt  
agnese.bissi@physics.uu.se \\ \hskip 42pt  
xinan.zhou@ucas.ac.cn}

\abstract{We initiate the study of one-loop gluon amplitudes in AdS space. These amplitudes were recently computed at tree level for a variety of backgrounds of the form $AdS_{d+1} \times S^3$. For concreteness, we compute the one-loop correction to the massless gluon amplitude on $AdS_5\times S^3$, which corresponds to the four-point correlator of the flavor current multiplet in the dual 4d $\mathcal{N}=2$ SCFT. This requires solving a mixing problem that involves tree-level amplitudes of arbitrarily massive Kaluza-Klein modes. The final answer has the same color structure as in flat space but the dependence on Mandelstam variables is more complicated, with logarithms replaced by polygamma functions. 
}

\begin{document}
\maketitle

\newpage

\section{Introduction}
Via the AdS/CFT correspondence, correlation functions in holographic CFTs are mapped to on-shell scattering amplitudes in AdS space. These fundamental observables encode important theoretical data, and play a central role in testing and exploiting the correspondence. Computing these holographic objects was once a notoriously difficult task because the complexity of perturbation theory is amplified by the spacetime curvature. Explicit results were available only in a handful special examples. In recent years, however, the application of bootstrap ideas has led to drastic simplifications in the calculation, prompting a great deal of new developments. A vast amount of results for amplitudes have now been obtained in a variety of supergravity theories, both at tree level \cite{Rastelli:2016nze,Rastelli:2017udc,Rastelli:2017ymc,Zhou:2017zaw,Caron-Huot:2018kta,Rastelli:2019gtj,Giusto:2019pxc,Goncalves:2019znr,Giusto:2020neo,Alday:2020lbp,Alday:2020dtb,Wen:2021lio} and at loop level \cite{Alday:2017xua,Aprile:2017bgs,Alday:2017vkk,Aprile:2017qoy,Alday:2018kkw,Aprile:2019rep,Alday:2019nin,Alday:2020tgi,Bissi:2020wtv,Bissi:2020woe,Alday:2021ymb}. The results exhibit a remarkable simplicity which is obscured by the diagrammatic expansion and is reminiscent of the situation in flat space. Moreover, these AdS supergravity amplitudes display many unexpected interesting structures, such as hidden conformal symmetries \cite{Caron-Huot:2018kta,Rastelli:2019gtj,Giusto:2020neo} and Parisi-Sourlas-like dimensional reduction \cite{Behan:2021pzk}. By contrast, amplitudes of super Yang-Mills theory in AdS so far have received  much less attention. Tree-level four-point amplitudes were computed only very recently in \cite{Alday:2021odx} for a class of backgrounds of the form $AdS_{d+1}\times S^3$, generalizing the result of an earlier work \cite{Zhou:2018ofp}.\footnote{For AdS gluon amplitudes in bosonic Yang-Mills theory with four or more points, recent works include \cite{Roehrig:2020kck,Armstrong:2020woi,Albayrak:2020fyp,Diwakar:2021juk}.}
Many aspects of AdS supergravity amplitudes have not been similarly explored in the gluon context. Nevertheless, there is a lot of incentives to further pursue the study of AdS gluon scattering. On the one hand, gluon amplitudes in general have simpler structures compared to graviton amplitudes.  This makes them easier targets to study and also more suitable arenas for new ideas. On the other hand, gluon and graviton amplitudes are known to be intimately connected in flat space by double copy relations.\footnote{See \cite{Bern:2019prr} for a recent review.} Gluon amplitudes are in a sense more fundamental, as graviton amplitudes can be constructed by ``squaring'' them. It is conceivable that some of these flat space properties extend to AdS space as well.

In this paper, we continue the investigation of gluon scattering in AdS and initiate the study of loop-level amplitudes. For concreteness, we focus on the cases where the hyperbolic space is $AdS_5$ and require the system to preserve eight Poincar\'e supercharges ({\it i.e.}, the dual SCFTs have 4d $\mathcal{N}=2$ superconformal symmetry). Such backgrounds arise in several top-down constructions. For example, we can consider a stack of D3-branes probing F-theory singularities \cite{Fayyazuddin:1998fb,Aharony:1998xz}, or a large number of D3-branes with a few probe D7-branes \cite{Karch:2002sh}. In the near horizon geometry, both classes of theories contain an $AdS_5\times S^3$ locus which carries localized degrees of freedom organized into an eight dimensional $\mathcal{N}=1$ vector multiplet. This vector multiplet transforms in the adjoint representation of the flavor group  $G_F$ which is a gauge group from the bulk perspective. The Kaluza-Klein reduction of this multiplet onto $AdS_5$ gives rise to an infinite tower of states which have at most Lorentz spin 1. These are the massless and massive gluons and their super partners. An important feature of these theories is that the coupling between the gluons with gravitons is parametrically smaller than the gluon self-coupling in the large central charge limit. Therefore, it is natural to decouple gravity and consider a spin-1 gauge theory on $AdS_5\times S^3$ which describes the leading order dynamics of the gluonic sector. Furthermore, we can also consider a consistent truncation of the gluon theory to its lowest Kaluza-Klein level (the massless sector). This gives rise to an $AdS_5$ toy model without the $S^3$ internal manifold.

\begin{figure}[h]
\centering
\includegraphics[width=0.8\textwidth]{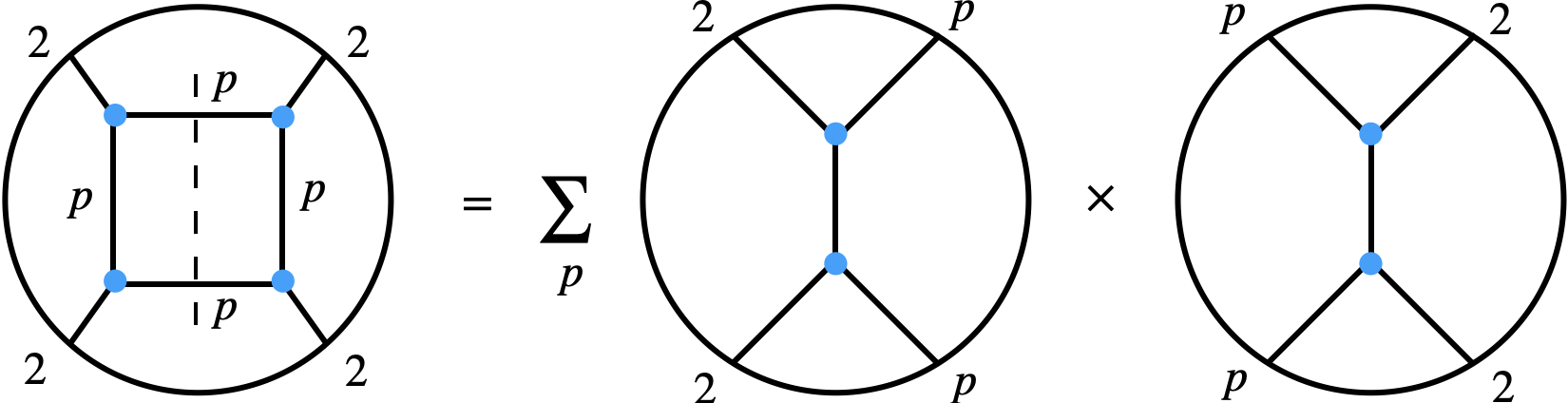}
\caption{The unitarity method in AdS. The $\langle 2222 \rangle$ one-loop super gluon amplitude in $AdS_5\times S^3$ can be constructed by ``gluing'' together all $\langle 22pp\rangle$ tree-level amplitudes. For the case of $AdS_5$ without internal manifold, there are no higher Kaluza-Klein modes and $p$ is restricted to 2.}
    \label{fig:um}
\end{figure}

Specifically, we will consider scattering amplitudes of {\it super gluons}. They are scalar super partners of the spin-1 gluons, and are the super primaries of the superconformal multiplets. The super gluons are labelled by an integer Kaluza-Klein level $k=2,3,\ldots$, which determines their squared masses in AdS to be $m^2=k(k-4)$. Furthermore, we will use the Mellin representation \cite{Mack:2009mi,Penedones:2010ue} to facilitate the computation. In this formalism AdS scattering amplitudes become Mellin amplitudes, which share a lot of similarities with flat space amplitudes. Thanks to supersymmetry, we can further express the Mellin amplitudes in terms of simpler {\it reduced} Mellin amplitudes. In flat space, this reduction is analogous to extracting a fermionic delta function from the super amplitudes. One of the main results of this paper is a closed form expression for the four-point one-loop reduced Mellin amplitude of $k_i=2$ super gluons, for any gauge group $G_F$. To obtain this result, our main technical tool is the AdS unitarity method  developed in \cite{Aharony:2016dwx}, which is illustrated in Figure \ref{fig:um}. Intuitively, the one-loop amplitude can be obtained by ``gluing'' together pairs of tree-level four-point amplitudes and summing over all the modes which run in the loop. From the figure, we can see that the necessary tree-level input for $AdS_5\times S^3$ is the $\langle22pp\rangle$ amplitudes for arbitrary values of $p$, which have already been computed in \cite{Alday:2021odx}. The gluing procedure will be made precise in the paper, and its realization in Mellin space turns out to be similar to the supergravity case \cite{Alday:2018kkw,Alday:2019nin}. However, a major difference is that the super gluons transform in the adjoint representation of the color group. This makes the amplitude contain multiple independent color structures which are reshuffled under crossing. These color structures add an extra layer of complication which is absent in the supergravity case. We will perform gluing in each color channel, which gives part of the answer. By further using crossing symmetry we will show that the one-loop reduced amplitude is uniquely fixed. However, this procedure does not make it clear whether the result should sensitively depend on the chosen gauge group. To answer this question, we find that the reduced Mellin amplitude can be rewritten in the following remarkably simple form
\begin{equation}
\widetilde{\mathcal{M}}(s,t)\sim \mathtt{d}_{st}\mathcal{B}_{st}(s,t)+\mathtt{d}_{su}\mathcal{B}_{su}(s,t)+\mathtt{d}_{tu}\mathcal{B}_{tu}(s,t)
\end{equation} 
where the three terms are related by crossing symmetry. The function $\mathcal{B}_{st}(s,t)$ is essentially a box diagram depicted in the LHS of Figure \ref{fig:um}, and has the form of an infinite sum of simultaneous simple poles 
\begin{equation}
\mathcal{B}_{st}(s,t)=\sum_{m,n=2}^\infty \frac{c_{mn}}{(s-2m)(t-2n)}
\end{equation}
where $c_{mn}$ are constants independent of the Mellin-Mandelstam variables. Interestingly, similar building block amplitudes also appeared in the supergravity one-loop reduced amplitudes  \cite{Alday:2018kkw,Alday:2019nin}. In this paper we give a closed form expression for $\mathcal{B}_{st}(s,t)$ in terms of special functions, see (\ref{AdSbox}). On the other hand, the factors $\mathtt{d}_{st}$, $\mathtt{d}_{su}$, $\mathtt{d}_{tu}$ are ``box diagrams'' in color space and are formed by contracting the structure constants of $G_F$ in the following way
\begin{equation}
\mathtt{d}_{st}= f^{JI_1K}f^{KI_2L}f^{LI_3M}f^{MI_4J}\;.
\end{equation}
Note that the dependence on the gauge group only enters the reduced amplitude through the structure constants $f^{IJK}$ (and also in the overall constant suppressed here). Therefore the answer above is in fact valid for any choice of the gauge group. One might also notice that the structure of the above result is reminiscent of the gluon one-loop amplitude in flat space. In fact, we can take the zero-curvature limit of the AdS amplitude and show that $\widetilde{\mathcal{M}}(s,t)$ reproduces exactly the 8d flat space amplitude. As another interesting example, we will also compute the super gluon one-loop amplitude in the $AdS_5$ toy model with no internal manifold. In applying the unitarity method, the Kaluza-Klein level $p$ is now restricted to 2 because the higher Kaluza-Klein modes have been excluded from the spectrum. It turns out that this modification does not alter the structure of the reduced Mellin amplitude but changes only the $c_{mn}$ coefficients. Taking the flat space limit, we find that the amplitude $\mathcal{B}_{st}(s,t)$ becomes the 5d one-loop box diagram instead. 

The rest of the paper is organized as follows. In Section \ref{Sec:prelim} we review the setup of the problem and the basic kinematics of four-point functions. In Section \ref{Sec:tree} we analyze the tree-level super gluon amplitudes and extract data from them. The one-loop computation was first performed for the simpler $AdS_5$ toy model in Section \ref{Sec:loopAdS5}  and then for the full-fledged $AdS_5\times S^3$ theory in Section \ref{Sec:loopAdS5S3}. In Section \ref{Sec:coupgravity}, we consider another contribution to the super gluon four-point function, which arises from coupling to gravity and is at the same order as the one-loop gluon contribution. We conclude with a brief discussion of future directions in Section \ref{Sec:disc}. Various technical details are relegated to the three appendices. In Appendix \ref{scfblocks} and \ref{App:Dfunctions} we collect some useful results for superconformal blocks and $D$-functions. In Appendix \ref{App:flatspacebox} we explain how to take the flat space limit of one-loop Mellin amplitudes and how they reduce to flat space box integrals.

\section{Preliminaries}\label{Sec:prelim}
\subsection{Setup}
In this paper, we study scattering amplitudes of super gluons in $AdS_5$. Such super gluons can arise in two basic setups. The first is to consider a stack of $N$ D3-branes probing an F-theory 7-brane singularity \cite{Fayyazuddin:1998fb,Aharony:1998xz}. The near horizon geometry has a metric identical to that of $AdS_5\times S^5$, except that one of the angular coordinates of $S^5$ has a changed periodicity. More precisely, we have
\begin{equation}
ds^2=d\theta^2+\sin^2\theta d\phi^2+\cos^2\theta d\Omega_3^2
\end{equation}
where $d\Omega_3^2$ is the metric of $S^3$, and $0\leq \theta \leq \frac{\pi}{2}$. The angular variable $\phi$ has a period of $2\pi(1-\nu/2)$, where $\nu$ takes values
\begin{equation}\label{nuvalues}
\nu=\frac{1}{3}\;,\; \frac{1}{2}\;,\;\frac{2}{3}\;,\;1\;,\;\frac{4}{3}\;,\;\frac{3}{2}\;,\;\frac{5}{3}\;,
\end{equation}
depending on the type of the 7-brane singularity. The 7-brane is located at the slice $\theta=0$, which fills the $AdS_5$ and occupies an $S^3$ in the compact space. On this singular locus, there is a $7+1$ dimensional $\mathcal{N}=1$ vector multiplet which transforms in the adjoint representation of a gauge group $G_F$. Corresponding to the values of $\nu$ in (\ref{nuvalues}), $G_F$ is given by
\begin{equation}
G_F=U(1)\;,\; SU(2)\;,\; SU(3)\;,\; SO(8)\;,\; E_6\;,\; E_7\;,\; E_8.
\end{equation}
The Kaluza-Klein reduction of the $\mathcal{N}=1$ vector multiplet gives rise to an infinite tower of states which are organized into different  $\frac{1}{2}$-BPS super multiplets of the 4d $\mathcal{N}=2$ superconformal algebra. We refer to the superprimaries of these $\frac{1}{2}$-BPS super multiplets as the {\it super gluons}. The $SU(2)_R$ R-symmetry of the 4d $\mathcal{N}=2$ superconformal group is identified with one of the $SU(2)$ factors of the $S^3$ isometry group $SO(4)\simeq SU(2)_R\times SU(2)_L$, while the other $SU(2)_L$ factor is a global symmetry. On the other hand, gravity is not restricted to this subspace. Instead, it propagates in the full ten dimensional space. The Kaluza-Klein reduction of the ten dimensional supergravity multiplet gives rise to an infinite tower of super graviton multiplets in $AdS_5$. An interesting feature of such systems is that at large $N$ the self-interactions of the super gluons are parametrically larger than their couplings with the super gravitons. More precisely, the cubic couplings of super gluons scale as $1/\sqrt{N}$, while the couplings involving two super gluons and one super graviton scale as $1/N$. Therefore, in $1/N$ expansion the super gluon four-point functions have the form 
\begin{equation}
G_{4\mathrm{-gluon}}=G_{\rm disc}+\frac{1}{N}G_{\text{gluon tree}}+\frac{1}{N^2}(G_{\text{gluon 1-loop}}+G_{\text{graviton tree}})+\ldots\;,
\end{equation}
where $G_{\rm disc}$ is the disconnected contribution, $G_{\text{gluon tree}}$ and $G_{\text{graviton tree}}$ respectively come from tree-level gluon and graviton exchanges, and $G_{\text{gluon 1-loop}}$ is the gluon one-loop contribution. 

The second setup is to consider a stack of $N_F$ D7-branes wrapping an $AdS_5\times S^3$ subspace in the $AdS_5\times S^5$ near horizon geometry of $N$ D3-branes \cite{Karch:2002sh}. In the limit $N_F\ll N$, the probe D7-branes do not back-react the $AdS_5\times S^5$ geometry. The theory is conformally invariant and preserves a 4d $\mathcal{N}=2$ superconformal symmetry. The low energy degrees of freedom on the D7-branes are again described by an eight dimensional $\mathcal{N}=1$ vector multiplet which transforms in the adjoint representation of a gauge group $SU(N_F)$. Reducing this multiplet onto $AdS_5$ leads to the same spectrum of super gluons. Just as in the previous setup, there is a separation of scales in the super gluon self-couplings and their couplings to the super gravitons at large $N$. These couplings also have the same $1/N$ scaling. Therefore at $\mathcal{O}(1/N)$, the super gluon four-point functions consist of only tree-level exchange contributions of super gluons. As was shown in \cite{Alday:2021odx},  the tree-level gluon exchange amplitudes are completely determined by the spectrum and superconformal symmetry, and have the same form for all choices of gauge groups. Therefore the super gluon tree-level amplitudes at $\mathcal{O}(1/N)$  are the same in both setups up to an overall constant. 

Finally, we can also consistently truncate the $AdS_5\times S^3$ super gluon theory to the lowest Kaluza-Klein mode, and obtain an $AdS_5$ theory without an internal $S^3$. The tree-level super gluon exchange amplitude of the lowest Kaluza-Klein mode remains the same after the truncation. We will use this truncated theory as a toy model when studying the super gluon amplitudes at one loop.

\subsection{Super gluon four-point functions}
The above mentioned super gluons are dual to scalar superprimaries $\mathcal{O}_k^{I;a_1\ldots a_k;\bar{a}_1\ldots \bar{a}_{k-2}}$ of the $\frac{1}{2}$-BPS multiplets with $k=2,3,\ldots$. Here $I=1,\ldots, {\rm dim}(G_F)$ is the color index, in the adjoint representation of the flavor group $G_F$. The $SU(2)$ indices $a_i=1,2$ and $\bar{a}_i=1,2$ are respectively for $SU(2)_R$ and $SU(2)_L$. The former is the R-symmetry group of the 4d $\mathcal{N}=2$ superconformal algebra, while the latter is an additional global symmetry. These operators have spin $\frac{k}{2}$ under $SU(2)_R$ and spin $\frac{k}{2}-1$ under $SU(2)_L$. Because the operators are $\frac{1}{2}$-BPS, their conformal dimensions are determined by the $SU(2)_R$ spins to be $k$.

It is convenient to keep track of the $SU(2)$ indices by contracting them with auxiliary two-component spinors $v^a$ and $\bar{v}^{\bar{a}}$
\begin{equation}
\mathcal{O}_k^{I}(x;v,\bar{v})=\mathcal{O}_k^{I;a_1\ldots a_k;\bar{a}_1\ldots \bar{a}_{k-2}}v_{a_1}\ldots v_{a_k}\bar{v}_{\bar{a}_1}\ldots \bar{v}_{\bar{a}_{k-2}}
\end{equation}
where indices are lowered with $\epsilon$ tensors $v_a=v^b\epsilon_{ab}$, $\bar{v}_{\bar{a}}=\bar{v}^{\bar{b}}\epsilon_{\bar{a}\bar{b}}$. We can then consider four-point correlators of such operators
\begin{equation}
G_{k_1k_2k_3k_4}^{I_1I_2I_3I_4}(x_i;v_i,\bar{v}_i)=\langle \mathcal{O}_{k_1}^{I_1}(x_1;v_1,\bar{v}_1)\mathcal{O}_{k_2}^{I_2}(x_2;v_2,\bar{v}_2)\mathcal{O}_{k_3}^{I_3}(x_3;v_3,\bar{v}_3)\mathcal{O}_{k_4}^{I_4}(x_4;v_4,\bar{v}_4)\rangle\;,
\end{equation}
which are functions of both the spacetime and the internal coordinates. For the purposes of this paper it is enough to consider pairwise equal external weights $k_1=k_2=p$ and $k_3=k_4=q$, and will focus on such correlators from now on. For a discussion on the kinematics of correlators with most general weights, see \cite{Alday:2021odx}. Exploiting conformal symmetry and the two $SU(2)$ symmetries, we can write the correlators as functions of four cross ratios
\begin{equation} \label{fourpoint}
G_{ppqq}^{I_1I_2I_3I_4}=\frac{(v_1\cdot v_2)^p(v_3\cdot v_4)^q(\bar{v}_1\cdot \bar{v}_2)^{p-2}(\bar{v}_3\cdot \bar{v}_4)^{q-2}}{x_{12}^{2p}x_{34}^{2q}}\mathcal{G}_{ppqq}^{I_1I_2I_3I_4}(U,V;\alpha,\beta)\;.
\end{equation}
Here $x_{ij}=x_i-x_j$, $v_i\cdot v_j=v_i^av_j^b\epsilon_{ab}$, $\bar{v}_i \cdot \bar{v}_j=\bar{v}_i^{\bar{a}}\bar{v}_j^{\bar{b}}\epsilon_{\bar{a}\bar{b}}$, and the cross ratios are defined as 
\begin{equation}
U=\frac{x_{12}^2x_{34}^2}{x_{13}^2x_{24}^2}\;,\quad V=\frac{x_{14}^2x_{23}^2}{x_{13}^2x_{24}^2}\;,
\end{equation}
\begin{equation}
\alpha=\frac{(v_1\cdot v_3)(v_2\cdot v_4)}{(v_1\cdot v_2)(v_3\cdot v_4)}\;,\quad \beta=\frac{(\bar{v}_1\cdot \bar{v}_3)(\bar{v}_2\cdot \bar{v}_4)}{(\bar{v}_1\cdot \bar{v}_2)(\bar{v}_3\cdot \bar{v}_4)}\;.
\end{equation}
It is easy to see that $\mathcal{G}_{ppqq}^{I_1I_2I_3I_4}$ is a polynomial of degree $\min\{p,q\}$ in $\alpha$, and a polynomial of degree $\min\{p,q\}-2$ in $\beta$. 

We have so far only exploited the bosonic symmetries. The fermionic generators in the superconformal group generate extra constraints known as the superconformal Ward identities \cite{Nirschl:2004pa}. To write down these identities, we make a change of variables
\begin{equation}
U=z\bar{z}\;,\quad V=(1-z)(1-\bar{z})\;.
\end{equation}
Then superconformal Ward identities read  
\begin{equation}\label{scfWardidposi}
\begin{split}
&(z\partial_z-\alpha\partial_\alpha)\mathcal{G}_{ppqq}^{I_1I_2I_3I_4}(z,\bar{z};\alpha,\beta)\big|_{\alpha=1/z}=0\;,\\
&(\bar{z}\partial_{\bar{z}}-\alpha\partial_\alpha)\mathcal{G}_{ppqq}^{I_1I_2I_3I_4}(z,\bar{z};\alpha,\beta)\big|_{\alpha=1/\bar{z}}=0\;.
\end{split}
\end{equation}
These constraints can be solved and lead to solutions of the form   
\begin{equation}\label{solscfWI1}
\mathcal{G}_{ppqq}^{I_1I_2I_3I_4}=\mathcal{G}_{0,ppqq}^{I_1I_2I_3I_4}+\mathcal{R}\,\mathcal{H}_{ppqq}^{I_1I_2I_3I_4}
\end{equation}
where $\mathcal{R}$ is a factor determined by superconformal symmetry
\begin{equation}\label{calR}
\mathcal{R}=(1-z\alpha)(1-\bar{z}\alpha)\;,
\end{equation}
and $\mathcal{G}_{0,ppqq}^{I_1I_2I_3I_4}$ is a function that contains coupling-independent information and becomes holomorphic (or anti-holomorphic) when setting $\alpha=1/\bar{z}$ (or $\alpha=1/z$). The function $\mathcal{H}$ is called the {\it reduced} correlator and contains all the dynamical information. It is always possible to find $\mathcal{G}_0$ and $\mathcal{H}$ such that each of them separately enjoys Bose symmetry\footnote{In the case of identical operators, Bose symmetry is just crossing symmetry.}. More precisely, we can restore the stripped kinematic factor and rewrite (\ref{solscfWI1}) as 
\begin{equation}\label{GG0andH}
G_{ppqq}^{I_1I_2I_3I_4}=G_{0,ppqq}^{I_1I_2I_3I_4}+{\rm R}\, H_{ppqq}^{I_1I_2I_3I_4}
\end{equation}
where 
\begin{equation}
{\rm R}=(v_1\cdot v_2)^2(v_3\cdot v_4)^2x_{13}^2x_{24}^2\mathcal{R}
\end{equation}
is crossing symmetric. The protected part $G_0$ has the original conformal dimensions and $SU(2)_R\times SU(2)_L$ weights. On the other hand, $H$ can be viewed as a four-point correlator of operators with shifted conformal dimensions $k_i\to k_i+1$ and $SU(2)_R$ R-symmetry spins $\frac{k_i}{2}\to \frac{k_i}{2}-1$. These shifts are due to the nontrivial weights carried by the factor ${\rm R}$ under conformal and $SU(2)_R$ transformations. Note that in particular the reduced correlator with $k_i=2$ is a singlet under both $SU(2)_R$ and $SU(2)_L$.

A convenient language to discuss holographic correlators is the Mellin formalism \cite{Mack:2009mi,Penedones:2010ue}. In this representation, holographic correlators have simple analytic structure which closely resembles that of flat space scattering amplitudes. The Mellin amplitudes are defined by\footnote{Here we have restricted the discussion to $\langle ppqq\rangle$ correlators. For the most general weight configurations, we refer the reader to \cite{Alday:2021odx} for details.}
\begin{equation}
\mathcal{G}_{ppqq}^{I_1I_2I_3I_4}=\int_{-i\infty}^{i\infty}\frac{dsdt}{(4\pi i)^2}U^{\frac{s}{2}}V^{\frac{t-p-q}{2}}\mathcal{M}_{ppqq}\Gamma[\frac{2p-s}{2}]\Gamma[\frac{2q-s}{2}]\Gamma^2[\frac{p+q-t}{2}]\Gamma^2[\frac{p+q-u}{2}]
\end{equation}
where $s+t+u=\sum_i k_i=2(p+q)$. However, we can also define a {\it reduced} Mellin amplitude from the reduced correlator 
\begin{equation}
\mathcal{H}_{ppqq}^{I_1I_2I_3I_4}=\int_{-i\infty}^{i\infty}\frac{dsdt}{(4\pi i)^2}U^{\frac{s}{2}}V^{\frac{t-p-q}{2}}\widetilde{\mathcal{M}}_{ppqq}\Gamma[\frac{2p-s}{2}]\Gamma[\frac{2q-s}{2}]\Gamma^2[\frac{p+q-t}{2}]\Gamma^2[\frac{p+q-\tilde{u}}{2}]\;.
\end{equation}
Note that the $u$ variable is replaced by its shifted version $\tilde{u}=u-2$, and this shift is necessary due to the nontrivial conformal weights of ${\rm R}$. Bose symmetry then acts on the Mellin amplitude $\mathcal{M}$ by permuting $s$, $t$, $u$, in addition to permuting the quantum numbers of each operator. However, acting on the reduced Mellin amplitudes Bose symmetry permutes instead $s$, $t$, $\tilde{u}$. Thanks to the superconformal Ward identities, the reduced amplitude determines the full amplitude.  The protected part $\mathcal{G}_0$ for holographic correlators turn out to take a form resembling Wick contractions in mean field theories. As was argued in \cite{Rastelli:2017udc}, such functions do not contribute to Mellin amplitudes. By translating (\ref{solscfWI1}) into Mellin space, we find 
\begin{equation}
\mathcal{M}_{ppqq}=\mathbb{R}\circ\widetilde{\mathcal{M}}_{ppqq}
\end{equation}
where $\mathbb{R}$ is a difference operator descending from the factor $\mathcal{R}$. The explicit action of this operator can be found in \cite{Alday:2021odx} but is not needed in this paper. When computing one-loop correlators, we will focus on the reduced Mellin amplitudes which have already taken into account superconformal symmetry.

\subsection{Flavor symmetry}\label{Subsec:flavorsymm}
As noted in the previous subsection, the four-point correlators of super gluons carry flavor indices. This adds an extra layer of complication on top of the superconformal kinematics discussed above. Below we give a detailed discussion about how to deal with the flavor structures. 

A concrete way to discuss these structures is to decompose the correlator into different flavor channels that appear in the tensor product of adjoint representations in the s-channel
\begin{equation}
G^{I_1I_2I_3I_4}=\sum_{a\in {\rm adj}\otimes {\rm adj}} {\rm P}^{I_1I_2|I_3I_4}_a G_a\;.
\end{equation}
Here ${\rm P}^{I_1I_2|I_3I_4}_a$ are projectors which project the external indices to the irreducible representation $a$ appearing in ${\rm adj}\otimes {\rm adj}$, and form a complete basis for flavor structures. For example, the projectors associated with exchanging the identity and adjoint representations are given by
\begin{equation}\label{P1Padj}
{\rm P}^{I_1I_2|I_3I_4}_{\bf 1}=\frac{1}{{\rm dim}(G_F)}\delta^{I_1I_2}\delta^{I_3I_4}\;,\quad {\rm P}^{I_1I_2|I_3I_4}_{\rm adj}=\frac{1}{\psi^2 h^\vee}f^{I_1I_2I_5}f^{I_5I_3I_4}
\end{equation} 
where $h^\vee$ is the dual Coxeter number, $\psi^2$ is the length squared of the longest root, and $f^{IJK}$ are the structure constants of the flavor group $G_F$. The projectors satisfy symmetry properties 
\begin{equation}
{\rm P}^{I_1I_2|I_3I_4}_a=(-1)^{{\rm R}_a}{\rm P}^{I_2I_1|I_3I_4}_a\;,\quad {\rm P}^{I_1I_2|I_3I_4}_a={\rm P}^{I_3I_4|I_1I_2}_a
\end{equation}
where ${\rm R}_a$ is the parity of the exchanged representation, {\it i.e.}, ${\rm R}_a=0$ for symmetric representations and ${\rm R}_a=1$ for antisymmetric representations. They are also idempotent 
\begin{equation}\label{idempo}
{\rm P}^{I_1I_2|I_3I_4}_a{\rm P}^{I_4I_3|I_5I_6}_b=\delta_{ab}{\rm P}^{I_1I_2|I_5I_6}_a\;.
\end{equation}
Contracting the external indices gives a delta function for the internal representations 
\begin{equation}\label{Pdimform}
{\rm P}^{I_1I_2|I_3I_4}_a{\rm P}^{I_1I_2|I_3I_4}_b=\delta_{ab}{\rm dim}({\rm R}_a)\;.
\end{equation}
We can use this property to compute the dimension of representation $a$. 

\begin{figure}[h]
\centering
\includegraphics[width=0.65\textwidth]{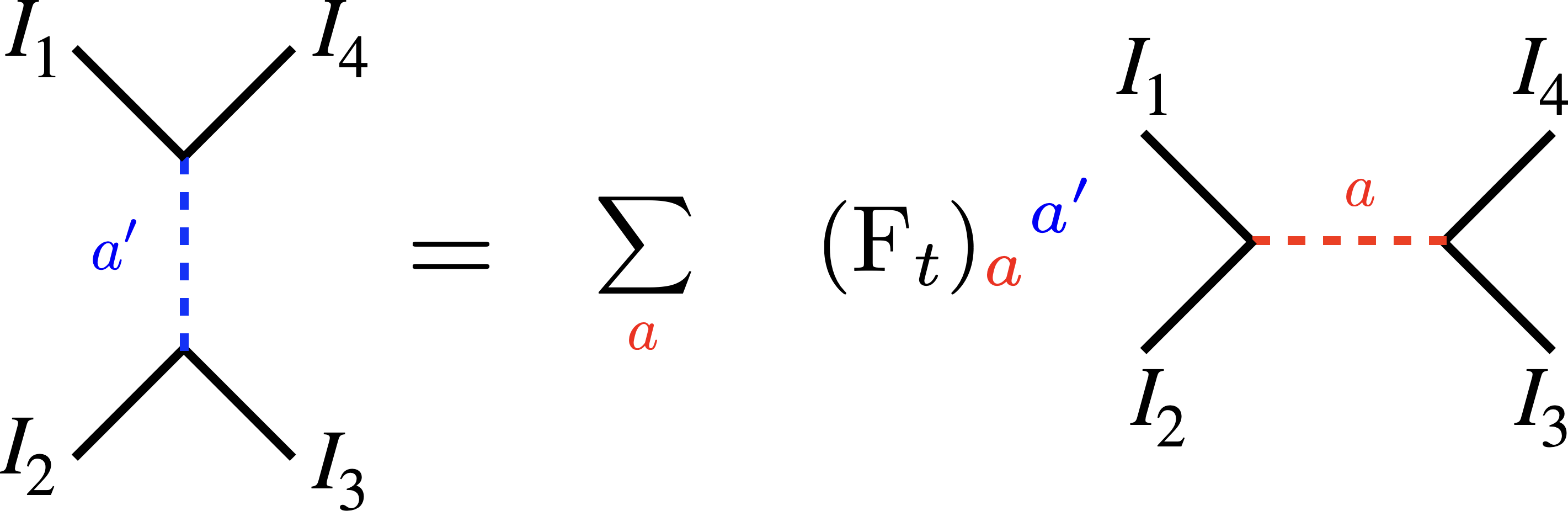}
\caption{Decomposition of a t-channel projector into s-channel projectors. The projectors in the two channels are related by the flavor crossing matrix ${\rm F}_t$. Note that here the vertices are oriented, and we read the lines around a vertex anti-clock-wisely.}
    \label{fig:Ft}
\end{figure}

We will also encounter the problem of decomposing exchanged flavor representation in a different channel. This amounts to performing a change of basis. Consider exchanging the representation $a'$ in the t-channel, which is captured by the t-channel projector ${\rm P}^{I_3I_2|I_1I_4}_{a'}$. We can decompose it into various s-channel representations as
\begin{equation}
{\rm P}^{I_3I_2|I_1I_4}_{a'}=\sum_a \mu_a {\rm P}^{I_1I_2|I_3I_4}_a\;,
\end{equation}
using the completeness of the basis. To get the overlap coefficients, we use (\ref{Pdimform}) and get \begin{equation}
\mu_a=\frac{1}{{\rm dim}({\rm R}_a)}{\rm P}^{I_3I_2|I_1I_4}_{a'} {\rm P}^{I_1I_2|I_3I_4}_a \;.
\end{equation}
Let us define the t-channel {\it flavor crossing matrix} 
\begin{equation}\label{defFt}
({\rm F}_t)_{a}{}^{a'}\equiv \frac{1}{{\rm dim}({\rm R}_a)}{\rm P}^{I_3I_2|I_1I_4}_{a} {\rm P}^{I_1I_2|I_3I_4}_{a'}\;,
\end{equation}
then we can express the overlap coefficients as 
\begin{equation}
\mu_a=({\rm F}_t)_{a}{}^{a'}\;.
\end{equation}
Diagrammatically, this is represented by Figure \ref{fig:Ft}.

Therefore, it follows that a t-channel ``exchange'' can be expressed in the s-channel as 
\begin{equation}\label{ttos}
\sum_{a'\in ({\rm adj}\otimes {\rm adj})_t} {\rm P}^{I_3I_2|I_1I_4}_{a'}\!\!\!\!\!\!\!\! \underbrace{H^{(t)}_{a'}}_{\text{t-channel coefficients}}\!\!\!\!=\sum_{a\in ({\rm adj}\otimes {\rm adj})_s} {\rm P}^{I_1I_2|I_3I_4}_a \underbrace{\sum_{a'\in ({\rm adj}\otimes {\rm adj})_t}  ({\rm F}_t)_{a}{}^{a'} H^{(t)}_{a'}}_{\text{s-channel coefficients}}\;.
\end{equation}
Similarly, for the u-channel we define the crossing matrix 
\begin{equation}
({\rm F}_u)_{a}{}^{a'}\equiv \frac{1}{{\rm dim}({\rm R}_a)}{\rm P}^{I_4I_2|I_3I_1}_{a} {\rm P}^{I_1I_2|I_3I_4}_{a'}\;.
\end{equation}
Then the u-channel decomposition can be rewritten in the s-channel as 
\begin{equation}\label{utos}
\sum_{a'\in ({\rm adj}\otimes {\rm adj})_u} {\rm P}^{I_4I_2|I_3I_1}_{a'}\!\!\!\!\!\!\!\! \underbrace{H^{(u)}_{a'}}_{\text{u-channel coefficients}}\!\!\!\!=\sum_{a\in ({\rm adj}\otimes {\rm adj})_s} {\rm P}^{I_1I_2|I_3I_4}_a \underbrace{\sum_{a'\in ({\rm adj}\otimes {\rm adj})_u}  ({\rm F}_u)_{a}{}^{a'} H^{(u)}_{a'}}_{\text{s-channel coefficients}}\;.
\end{equation}
The elements of these flavor crossing matrices can be expressed in terms of the $3j$ and $6j$ symbols of the flavor group. We will not go into the details of computing the matrix elements.  Instead we will refer the reader to the book \cite{Cvitanovic:2008zz} for techniques of performing such computations, and Table 6 of \cite{Chang:2017cdx} for a list of examples. 

Finally, in this paper we will use the group $E_8$ as a nontrivial explicit example to demonstrate various computations -- although all our results can also be phrased abstractly without reference to the explicit crossing matrices, as we will show. We record here the crossing matrices of $E_8$ \cite{Chang:2017xmr}
\begin{equation} 
{\rm F}_t=\left(
\begin{array}{ccccc}
 \frac{1}{248} & \frac{125}{8} & \frac{3375}{31} & 1 & \frac{245}{2} \\
 \frac{1}{248} & -\frac{3}{8} & \frac{27}{31} & \frac{1}{5} & -\frac{7}{10} \\
 \frac{1}{248} & \frac{1}{8} & \frac{23}{62} & -\frac{1}{30} & -\frac{7}{15} \\
 \frac{1}{248} & \frac{25}{8} & -\frac{225}{62} & \frac{1}{2} & 0 \\
 \frac{1}{248} & -\frac{5}{56} & -\frac{90}{217} & 0 & \frac{1}{2} \\
\end{array}
\right)\;,\quad {\rm F}_u=\left(
\begin{array}{ccccc}
 \frac{1}{248} & \frac{125}{8} & \frac{3375}{31} & -1 & -\frac{245}{2} \\
 \frac{1}{248} & -\frac{3}{8} & \frac{27}{31} & -\frac{1}{5} & \frac{7}{10} \\
 \frac{1}{248} & \frac{1}{8} & \frac{23}{62} & \frac{1}{30} & \frac{7}{15} \\
 -\frac{1}{248} & -\frac{25}{8} & \frac{225}{62} & \frac{1}{2} & 0 \\
 -\frac{1}{248} & \frac{5}{56} & \frac{90}{217} & 0 & \frac{1}{2} \\
\end{array}
\right)\;
\end{equation}
where from top to bottom along each column  (or from left to right in  each row) the representations are $a=\bf{1},\bf{3875},\bf{27000},\bf{248}\,({\rm adj}), \bf{30380}$. Note that the first three representations are symmetric, while the last two are anti-symmetric.

\subsection{Superconformal block decomposition}\label{sec:scfblocks}
To be able to reconstruct the one-loop correlator, we need to decompose correlators into superconformal blocks. In order to understand the contribution of each superconformal multiplet, we study the fusion rules of the operators $\mathcal{O}$ appearing in our four-point function. 
Operators in $\mathcal{N}=2$ SCFTs are organized into different superconformal multiplets which have been classified in \cite{Nirschl:2004pa}. Here we will encounter three types of them: $\hat{\mathcal{B}}_{R}$, $\hat{\mathcal{C}}_{R,(\frac{\ell}{2},\frac{\ell}{2})}$ and $\mathcal{A}^\Delta_{R,(\frac{\ell}{2},\frac{\ell}{2})}$.\footnote{See Appendix A of \cite{Beem:2014zpa} for a summary of the unitary representations and different notations used in the literature.} The $\hat{\mathcal{B}}$ multiplets are $\frac{1}{2}$-BPS (known as short multiplets), and the conformal dimensions of the super primaries as fixed by the $SU(2)_R$ spin $R$ is $\frac{R}{2}$. The super gluons belong to this type of multiplets and are the corresponding super primaries. The $\hat{\mathcal{C}}$ multiplets are also protected, and are known as semi-short. The  super primary has $SU(2)_R$ spin $\frac{R}{2}$, Lorentz spin $\ell$ and dimension $2+2R+\ell$. The multiplets $\mathcal{A}$ are long multiplets, and the conformal dimensions are not protected by supersymmetry. These are the only three types of multiplets which appear in the fusion rules of two $\hat{\mathcal{B}}$ \cite{Nirschl:2004pa}
\begin{equation}
\hat{\mathcal{B}}_{\frac{k_1}{2}}\times \hat{\mathcal{B}}_{\frac{k_2}{2}}\simeq \bigoplus_{p=\frac{k_2}{2}-\frac{k_1}{2}}^{\frac{k_1}{2}+\frac{k_2}{2}}\hat{\mathcal{B}}_{p}+\bigoplus_{\ell\geq 0} \left(\bigoplus_{p=\frac{k_2}{2}-\frac{k_1}{2}}^{\frac{k_2}{2}+\frac{k_1}{2}-1}\hat{\mathcal{C}}_{p,(\frac{\ell}{2},\frac{\ell}{2})}+\bigoplus_{p=\frac{k_2}{2}-\frac{k_1}{2}}^{\frac{k_2}{2}+\frac{k_1}{2}-2}\mathcal{A}^\Delta_{p,(\frac{\ell}{2},\frac{\ell}{2})}\right)\;.
\end{equation}
The exchanged multiplets in the four-point function are those residing in the overlap of $\hat{\mathcal{B}}_{\frac{k_1}{2}}\times \hat{\mathcal{B}}_{\frac{k_2}{2}}$ and $\hat{\mathcal{B}}_{\frac{k_3}{2}}\times \hat{\mathcal{B}}_{\frac{k_4}{2}}$.
This is the most general fusion rules, but we will only need a less general setup to study the four point correlator \eqref{fourpoint}. In particular, it is more convenient to write the solution to the superconformal Ward identities in a slightly different form \cite{Nirschl:2004pa}
\begin{equation}\label{solscfWIfK}
\mathcal{G}(U,V;\alpha)=\frac{(y-\bar{x})f(\bar{x})-(y-x)f(x)}{x-\bar{x}}+(y-x)(y-\bar{x})\mathcal{K}(U,V;y)
\end{equation}
where we defined the new cross ratios
\begin{equation}
y=2\alpha-1\;,\quad x=\frac{2}{z}-1\;,\quad \bar{x}=\frac{2}{\bar{z}}-1\;.
\end{equation}
We have also temporarily suppressed the flavor indices in \eqref{fourpoint}, as well as the $SU(2)_L$ dependence, to lighten the notation. These extra structures are independent of the superconformal properties, and therefore can be easily restored. Given a four-point correlator $\mathcal{G}$, we can first obtain $f(x)$ by setting $y=\bar{x}$. Then from (\ref{solscfWIfK}) we can easily determine $\mathcal{K}$. In contrast to the function $\mathcal{H}$ defined in (\ref{solscfWI1}), $\mathcal{K}$ alone is {\it not} invariant under Bose symmetry. Instead, there is an inhomogeneous term coming from the single-variable function $f$ under crossing. 

The contribution of exchanging a superconformal multiplet to the four-point function is captured by  superconformal blocks, which are linear combinations of bosonic conformal blocks of the super primary and super descendants of the superconformal multiplet. They were obtained in \cite{Nirschl:2004pa} and are recorded in Appendix \ref{scfblocks}. In practice, it is most convenient to present them in terms of contributions to $f$ and $\mathcal{K}$ in (\ref{solscfWIfK}). Furthermore, the function $\mathcal{K}$ can be decomposed into different $SU(2)_R$ channels. For computing the one-loop amplitude of $\langle 2222\rangle$, only the R-symmetry singlet contributions are relevant.

To write down these contributions, let us first introduce the bosonic conformal blocks
\begin{equation}
G_{\Delta,\ell}(z,\bar{z})=\frac{(-\frac{1}{2})^{\ell}z\bar{z}}{z-\bar{z}}\big(k_{\frac{\Delta+\ell}{2}}(z)k_{\frac{\Delta-\ell-2}{2}}(\bar{z})-k_{\frac{\Delta+\ell}{2}}(\bar{z})k_{\frac{\Delta-\ell-2}{2}}(z)\big)
\end{equation}
where 
\begin{equation}
k_h(z)=z^h{}_2F_1(h,h;2h;z)\;,
\end{equation}
and this expression is valid in any four-point function with $\Delta_1=\Delta_2$, $\Delta_3=\Delta_4$. The long super multiplets $\mathcal{A}^\Delta_{R,(\frac{\ell}{2},\frac{\ell}{2})}$ do not contribute to $f(x)$. Moreover, only the long multiplets with $R=0$ contribute to the singlet channel of $\mathcal{K}$ by\footnote{We introduced a factor of $\frac{1}{4}$ such that when we expand the superconformal block into bosonic conformal blocks the conformal block of the super primary appears with a unit coefficient.}
\begin{equation}\label{longK}
\mathcal{K}^{\mathcal{A}}_{\Delta,R,\ell}\big|_{\rm sing}=\frac{1}{4}\delta_{R,0}G_{\Delta+2,\ell}(z,\bar{z})\;.
\end{equation}
The semi-short multiplets $\hat{\mathcal{C}}_{R,(\frac{\ell}{2},\frac{\ell}{2})}$ contribute
\begin{equation}\label{fKC}
\begin{split}
&f^{\mathcal{C}}_{R,\ell}(x)=\frac{(-1)^{\ell -1} 2^{2 R-\ell } \Gamma \left(R+\frac{1}{2}\right)}{\sqrt{\pi } \Gamma (R+1)}k_{R+2+\ell}(z)\;,\\
& \mathcal{K}^{\mathcal{C}}_{R,\ell}\big|_{\rm sing}=\frac{(-1)^{R+1} 2^{3 R-2} \Gamma \left(R+\frac{1}{2}\right)}{\sqrt{\pi } \Gamma (R+1)}G_{R+\ell+4,R+\ell}(z,\bar{z})\;.
\end{split}
\end{equation}
Finally, the short multiplets $\hat{\mathcal{B}}_R$ contribute 
\begin{equation}\label{fKB}
f^{\mathcal{B}}_{R}(x)=\frac{4^R \Gamma \left(R+\frac{1}{2}\right)}{\sqrt{\pi } \Gamma (R+1)}k_R(z)\;,\quad \mathcal{K}^{\mathcal{B}}_{R}\big|_{\rm sing} = \frac{(-1)^R 2^{3 R-4} \Gamma \left(R+\frac{1}{2}\right)}{\sqrt{\pi } \Gamma (R+1)}G_{R+2,R-2}(z,\bar{z})\;.
\end{equation}

From these expressions, it is easy to see that the contributions to $f$ from short and semi-short multiplets satisfy the following relations
\begin{equation}
f^{\mathcal{C}}_{R,\ell}=\frac{(-1)^{\ell +1} 2^{-3 \ell -4} \Gamma \left(R+\frac{1}{2}\right) \Gamma (R+\ell +3)}{\Gamma (R+1) \Gamma \left(R+\ell +\frac{5}{2}\right)} f^{\mathcal{B}}_{R+\ell+2}\;,
\end{equation}
\begin{equation}
f^{\mathcal{C}}_{R+a,\ell-a}=\frac{\left(-\frac{1}{8}\right)^{-a} \Gamma (R+1) \Gamma \left(a+R+\frac{1}{2}\right)}{\Gamma \left(R+\frac{1}{2}\right) \Gamma (a+R+1)} f^{\mathcal{C}}_{R,\ell}\;.
\end{equation}
These relations arise from the fact that short and semi-short multiplets can recombine and form long multiplets at unitarity bounds \cite{Nirschl:2004pa}. As a result, their contributions to $f$ cancel and they only contribute to the function $\mathcal{K}$. 

\vspace{0.5cm}

\noindent{\it Generalized free field theory}

\vspace{0.3cm}

As a simple application, let us consider the superconformal block decomposition of generalized free field (GFF) theory correlators. The result is also needed later for computing the one-loop amplitudes. 

The GFF correlators give the disconnected contribution of holographic correlators, which are leading in the $1/N$-expansion. We are interested in the correlators of four identical operators with dimension $k$ 
\begin{equation}
\mathcal{G}^{\rm GFF}_{kkkk}= \delta^{I_1I_2}\delta^{I_3I_4}+\delta^{I_1I_3}\delta^{I_2I_4}\mathcal{G}^{{\rm GFF},(u)}_{kkkk}+\delta^{I_1I_4}\delta^{I_2I_3}\mathcal{G}^{{\rm GFF},(t)}_{kkkk}\;.
\end{equation}
where 
\begin{equation}
\mathcal{G}^{{\rm GFF},(t)}_{kkkk}=(\alpha-1)^k(\beta-1)^{k-2}\frac{U^k}{V^k}\;,\quad \mathcal{G}^{{\rm GFF},(u)}_{kkkk}=\alpha^k\beta^{k-2}U^k\;.
\end{equation}
Let us look at the three independent flavor structures $\delta^{I_1I_2}\delta^{I_3I_4}$, $\delta^{I_1I_3}\delta^{I_2I_4}$, $\delta^{I_1I_4}\delta^{I_2I_3}$ separately. The coefficient functions of these structures can be straightforwardly rewritten in the form of (\ref{solscfWIfK}). Note that the first term $\delta^{I_1I_2}\delta^{I_3I_4}$ just corresponds to exchanging the identity operator in the s-channel. For the other two terms, we denote the $SU(2)_R\times SU(2)_L$ singlets of functions $\mathcal{K}$ from $\mathcal{G}^{{\rm GFF},(t,u)}_{kkkk}$  by  $\mathcal{K}^{(t,u)}_{k,{\rm sing}}$. We would like to know the OPE coefficients of long multiplets in these functions. To this end, we decompose $\mathcal{K}^{(t,u)}_{k,{\rm sing}}$ into conformal blocks and find 
\begin{eqnarray}
\mathcal{K}^{(t)}_{k,{\rm sing}}&=&\sum_{n=-k+1}^\infty \sum_{\ell=0}^\infty A_{n,\ell}^{(k)}G_{2k+2+2n+\ell,\ell}(z,\bar{z})\;,\\
\mathcal{K}^{(u)}_{k,{\rm sing}}&=&\sum_{n=-k+1}^\infty \sum_{\ell=0}^\infty (-1)^\ell A_{n,\ell}^{(k)}G_{2k+2+2n+\ell,\ell}(z,\bar{z})\;,
\end{eqnarray}
where 
\begin{equation}\label{A0kkkkGFF}
\begin{split}
 A_{n,\ell}^{(k)}=&\frac{\pi  (-1)^{\ell } (n+1)_{k-2} (k+n+1)_{k-2} (n+\ell +2)_{k-2} (k+n+\ell +2)_{k-2}}{2^{4 (k+n)+\ell }\Gamma (k)^4 \Gamma \left(k+n-\frac{1}{2}\right)}\\
& \times \frac{ (\ell +1) (2 (k+n)+\ell ) \Gamma (k+n)\Gamma (k+n+\ell +1)}{\Gamma \left(k+n+\ell +\frac{1}{2}\right)}\;.
\end{split}
\end{equation}
Note that due to the aforementioned multiplet recombination, there is an ambiguity in interpreting these conformal blocks. On the other hand, we note that recombination only happens at the unitarity bound. Since we are restricting ourselves to the $R=0$ sector, the conformal twist of the recombined long multiplets is $\tau=2$, {\it i.e.}, the conformal blocks $G_{4+\ell,\ell}(z,\bar{z})$.\footnote{In fact, we observe that the above $A_{n,\ell}^{(k)}$ coefficients with $n<0$ are all zero except for one value $n=-k+1$, which corresponds to $\tau=2$.} In the holographic limit to be discussed, which corresponds to turning on interactions for the generalized free field theory, we expect that all long multiplets are double-trace operators with minimal twist 4.\footnote{For the conformal block $G_{6+2n+\ell,\ell}(z,\bar{z})$, the associated operators have conformal twist $\tau=4+2n$ and are linear combinations of the double-trace operators $:\mathcal{O}_2\square^n\partial^\ell \mathcal{O}_2:$, $\mathcal{O}_3\square^{n-1}\partial^\ell \mathcal{O}_3:$, $ \ldots$, $:\mathcal{O}_{n+2}\partial^\ell \mathcal{O}_{n+2}:$.} Therefore, their coefficients are not affected by recombination and are simply given by (\ref{A0kkkkGFF}) with $n\geq 0$.

\section{Data from tree-level gluon amplitudes}\label{Sec:tree}
Let us proceed to tree level. The four-point super gluon Mellin amplitudes were computed in \cite{Alday:2021odx} for arbitrary external weights. To compute the $\langle 2222 \rangle$ correlator at one loop, we will only need the $\langle 22kk\rangle$ tree-level correlators.

\subsection{Correlators in position space}
To extract the CFT data, it is more convenient to rewrite the Mellin space result in position space. Let us first give the result written in the form of (\ref{GG0andH}), before outlining how the translation was performed. The protected parts for correlators $\langle 22kk\rangle$ with $k> 2$ are related to the $k=2$ case by
\begin{equation}
G_{0,22kk}=\frac{(v_3\cdot v_4)^{k-2}(\bar{v}_3\cdot \bar{v}_4)^{k-2}}{x_{34}^{2(k-2)}}G_{0,2222}\;,
\end{equation}
and the dynamical parts have the form 
\begin{equation}
H_{22kk}=(C_{2,2,2})^2\big(\mathtt{c}_sH_{22kk}^{(s)}+\mathtt{c}_tH_{22kk}^{(t)}+\mathtt{c}_uH_{22kk}^{(u)}\big)\;.
\end{equation} 
Here $\mathtt{c}_{s,t,u}$ are color structures 
\begin{equation}
\mathtt{c}_s=f^{I_1I_2J}f^{JI_3I_4}\;,\quad\quad \mathtt{c}_t=f^{I_1I_4J}f^{JI_2I_3}\;,\quad\quad \mathtt{c}_u=f^{I_1I_3J}f^{JI_4I_2}\;,
\end{equation}
which are related to the projectors defined in (\ref{Subsec:flavorsymm}) by 
\begin{equation}\label{ctoP}
\mathtt{c}_s=\psi^2h^\vee {\rm P}^{I_1I_2|I_3I_4}_{\rm adj}\;,\quad \mathtt{c}_t=-\psi^2h^\vee {\rm P}^{I_3I_2|I_1I_4}_{\rm adj}\;,\quad \mathtt{c}_u=-\psi^2h^\vee {\rm P}^{I_4I_2|I_3I_1}_{\rm adj}\;.
\end{equation}
Thanks to the Jacobi identity, they satisfy 
\begin{equation}\label{Jacobi}
\mathtt{c}_s+\mathtt{c}_t+\mathtt{c}_u=0\;.
\end{equation}
The function $G_{0,2222}$ has the form of Wick contractions
\begin{equation}\label{G02222}
\begin{split}
G_{0,2222}=&\frac{(v_1\cdot v_2)^2(v_3\cdot v_4)^2}{3x_{12}^2x_{34}^2}(C_{2,2,2})^2\bigg(\mathtt{c}_s\frac{(\alpha-1)x_{13}^2x_{24}^2+2\alpha x_{14}^2x_{23}^2+\alpha(1-\alpha)x_{12}^2x_{34}^2}{x_{13}^2x_{14}^2x_{23}^2x_{24}^2}\\
&\quad\quad\quad\quad\quad\quad\quad\quad\quad+\mathtt{c}_t \frac{(1-\alpha)x_{13}^2x_{24}^2+\alpha x_{14}^2x_{23}^2+2\alpha(1-\alpha)x_{12}^2x_{34}^2}{x_{13}^2x_{14}^2x_{23}^2x_{24}^2}\bigg)\;,
\end{split}
\end{equation}
while $H_{22kk}^{(s,t,u)}$ can be expressed in terms of $D$-functions (see Appendix \ref{App:Dfunctions} for their definition and properties)
\begin{eqnarray}\label{H22kk}
\nonumber H_{22kk}^{(s)}&=&\frac{2(k-1)k}{3\pi^2}\bigg(\frac{D_{1,2,k+1,k}}{x_{12}^2x_{14}^2}-\frac{D_{1,2,k,k+1}}{x_{12}^2x_{13}^2}\bigg)(v_3\cdot v_4)^{k-2}(\bar{v}_3\cdot \bar{v}_4)^{k-2}\;,\\
H_{22kk}^{(t)}&=&\frac{4(k-1)}{3\pi^2}\bigg(\frac{D_{1,3,k,k}}{x_{13}^2x_{14}^2}-\frac{k}{2}\frac{D_{1,2,k+1,k}}{x_{12}^2x_{14}^2}\bigg)(v_3\cdot v_4)^{k-2}(\bar{v}_3\cdot \bar{v}_4)^{k-2}\;,\\
\nonumber H_{22kk}^{(u)}&=&\frac{4(k-1)}{3\pi^2}\bigg(\frac{k}{2}\frac{D_{1,2,k,k+1}}{x_{12}^2x_{13}^2}-\frac{D_{1,3,k,k}}{x_{13}^2x_{14}^2}\bigg)(v_3\cdot v_4)^{k-2}(\bar{v}_3\cdot \bar{v}_4)^{k-2}\;.
\end{eqnarray}
Note that $H_{22kk}^{(s,t,u)}$ satisfy the relation
\begin{equation}
H_{22kk}^{(s)}+H_{22kk}^{(t)}+H_{22kk}^{(u)}=0\;,
\end{equation}
which parallels the Jacobi relation (\ref{Jacobi}). Finally, $C_{2,2,2}$ is related to the cubic coupling of the $AdS_5\times S^3$ SYM, and appears in the three-point function of $\mathcal{O}_2$ as 
\begin{equation}
\langle\mathcal{O}_2(x_1;v_1)\mathcal{O}_2(x_2;v_2)\mathcal{O}_2(x_3;v_3) \rangle=C_{2,2,2}\frac{(v_1\cdot v_2)(v_2\cdot v_3)(v_3\cdot v_1)}{x_{12}^2x_{13}^2x_{23}^2}\;.
\end{equation}
It can be expressed in terms of flavor central charge as\footnote{\label{fnsing}Here the flavor current two-point functions are 
\begin{equation}
\nonumber \langle \mathcal{J}_\mu^I(x)\mathcal{J}_\nu^J(0)\rangle=\frac{C_{\mathcal{J}}}{2\pi^2}\frac{\delta^{IJ}(\delta_{\mu\nu}-2\frac{x^\mu x^\nu}{x^2})}{x^6}\;.
\end{equation}
For the $\mathcal{N}=2$ SCFTs arising from D3-branes probing F-theory singularities, $C_{\mathcal{J}}$ is given by $C_{\mathcal{J}}=\frac{12}{2-\nu}N$, where $N$ is the number of D3-branes, and $\nu$ is determined by the type of the singularity according to (\ref{nuvalues}).} 
\begin{equation}
(C_{2,2,2})^2=\frac{6}{C_{\mathcal{J}}}\;.
\end{equation}

The translation from Mellin space to position space was done in two steps. One writes down an ansatz for $G$ in terms of exchange and contact Witten diagrams which are expressed in terms of $D$-functions using the method of \cite{DHoker:1999mqo}. The coefficients of these diagrams are found by matching the Mellin amplitudes. Similarly, one finds a position space expression for the reduced correlator $H$ in terms of $D$-functions. Then the protected piece $G_0$ can be solved from (\ref{GG0andH}), and can be simplified into the form (\ref{G02222}) after using a few $D$-function identities which are summarized in Appendix \ref{App:Dfunctions}.

Moreover, the $D$-functions in (\ref{H22kk}) satisfy a set of differential recursion relations which shifts their weights. These relations are collected in Appendix \ref{App:Dfunctions}, and allow us to relate $H_{22kk}^{(s,t,u)}$ with $k>2$ to $H_{2222}^{(s,t,u)}$ by differential operators. Let us define 
\begin{equation}
H_{22kk}^{(s,t,u)}(x_i,v_i,\bar{v}_i)=\frac{(v_3\cdot v_4)^{k-2}(\bar{v}_3\cdot \bar{v}_4)^{k-2}}{x_{12}^6x_{34}^{2+2k}}U\mathcal{H}_{22kk}^{(s,t,u)}(U,V)\;.
\end{equation}
Then using the formulae in Appendix \ref{App:Dfunctions}, one can show 
\begin{equation}\label{H22kkfromH2222}
U\mathcal{H}_{22kk}^{(s,t,u)}=\frac{1}{(k-2)!}\prod_{i=1}^{k-2} (i+2-U\partial_U)\big(U\mathcal{H}_{2222}^{(s,t,u)}\big)\;.
\end{equation}

\subsection{Extracting tree-level data}
We are interested in the tree-level anomalous dimensions which are encoded in the $\log U$ coefficients of the reduced correlators in the small $U$ expansion. Let us first focus on the case with $k=2$. Using the differential recursion relations, we find that the reduced correlator can be compactly written as
\begin{eqnarray}\label{Hstu2222}
\nonumber\mathcal{H}_{2222}^{(s)}=&&\frac{U^2}{3}\big(2\partial_U+(1+V)\partial_U\partial_V+U\partial_U^2\big)\Phi(z,\bar{z})\;,\\
\mathcal{H}_{2222}^{(t)}=&&-\frac{U^2}{3}\big(2\partial_V+V\partial_V^2+(1+U)\partial_U\partial_V\big)\Phi(z,\bar{z})\;,\\
\nonumber\mathcal{H}_{2222}^{(u)}=&&\frac{U^2}{3}\big(2\partial_V+V\partial_V^2-2\partial_U+(U-V)\partial_U\partial_V-U\partial_U^2\big)\Phi(z,\bar{z})
\end{eqnarray}
where $\Phi(z,\bar{z})=\bar{D}_{1,1,1,1}$ is the well known scalar one-loop box diagram. The $\log U$ coefficient of $\Phi(z,\bar{z})$ is given by 
\begin{equation}
\Psi(z,\bar{z})\equiv\Phi(z,\bar{z})\big|_{\log U}=\frac{\log(1-z)-\log(1-\bar{z})}{z-\bar{z}}\;.
\end{equation}
It is then straightforward to obtain the $\log U$ coefficients of $\mathcal{H}_{2222}^{(s,t,u)}$, simply by replacing $\Phi$ with $\Psi$ in (\ref{Hstu2222}), as differential operators acting on $\log U$ would lead to rational functions. Similarly, the $\log U$ coefficients of $\mathcal{H}_{22kk}^{(s,t,u)}$ are obtained by further acting on $\mathcal{H}_{2222}^{(s)}\big|_{\log U}$ with the differential operators in (\ref{H22kkfromH2222}). It is not difficult to decompose these $\log U$ coefficients into conformal blocks, and we find\footnote{We remind the reader the difference between $\mathcal{K}$ and $\mathcal{H}$. The factor $(x-y)(\bar{x}-y)$ in (\ref{solscfWIfK}) is $4U^{-1}\mathcal{R}$ where $\mathcal{R}$ was defined in (\ref{calR}). Therefore a long multiplet of which the super primary has dimension $\Delta$ and spin $\ell$ contributes to $\mathcal{H}$ by $U^{-1}G_{\Delta+2,\ell}$.} 
\begin{eqnarray}
\mathcal{H}_{22kk}^{(s)}\big|_{\log U}&=&\sum_{n,\ell}\left(\frac{1}{2}-\frac{(-1)^\ell}{2}\right)\omega_{n,\ell}^{(k)} U^{-1} G_{2k+2+2n+\ell,\ell}(z,\bar{z})\;,\\
\mathcal{H}_{22kk}^{(t)}\big|_{\log U}&=&\sum_{n,\ell}\left(\frac{1}{2}+(-1)^\ell\right)\omega_{n,\ell}^{(k)} U^{-1} G_{2k+2+2n+\ell,\ell}(z,\bar{z})\;,\\
\mathcal{H}_{22kk}^{(u)}\big|_{\log U}&=&-\sum_{n,\ell}\left(1+\frac{(-1)^\ell}{2}\right)\omega_{n,\ell}^{(k)} U^{-1} G_{2k+2+2n+\ell,\ell}(z,\bar{z})\;,
\end{eqnarray}
where 
\begin{equation}
\omega_{n,\ell}^{(k)}=\frac{\pi  (-1)^{k+1} (n+1)_{k-1} 2^{-4 k-4 n-\ell +3} \Gamma (2 k+n-1) \Gamma (k+n+\ell +1)}{3 \Gamma (k-1) \Gamma (k) \Gamma \left(k+n-\frac{1}{2}\right) \Gamma \left(k+n+\ell +\frac{1}{2}\right)}\;.
\end{equation}
The coefficients $\omega_{n,\ell}^{(k)}$ are proportional to the schematic averaged quantity $\langle a^{(0)}_{n,\ell}\gamma^{(1)}_{n,\ell} \rangle_a$ in each color channel
\begin{equation}
\frac{1}{2}\sum_{n,\ell} \langle a^{(0)}_{n,\ell}\gamma^{(1)}_{n,\ell} \rangle_a  U^{-1} G_{2k+2+2n+\ell,\ell}(z,\bar{z})={\rm P}_a^{I_1I_2|I_3I_4} \mathcal{H}_{22kk}\big|_{\log U}
\end{equation}
where $a^{(0)}_{n,\ell}$ are generalized free field theory OPE coefficients and $\gamma^{(1)}_{n,\ell}$ are the tree-level anomalous dimensions of the double-trace operators.


\section{Toy model: AdS$_5$ without internal S$^3$}\label{Sec:loopAdS5}
In this section, we first study one-loop amplitudes in a toy model where the $AdS_5$ super gluon theory does not have an internal $S^3$ manifold. This amounts to setting all $k>2$ fields to zero in the $AdS_5\times S^3$ theories, which consistently truncates the theory to the $k=2$ sector. As a result, there is no operator mixing in this model. All the double-trace operators are of the schematic form $:\mathcal{O}_2\square^n\partial^\ell\mathcal{O}_2:$, and there is a one-to-one correspondence between the conformal twist and $n$.\footnote{By contrast, the $AdS_5\times S^3$ theories allow, {\it e.g.}, operators of the form $:\mathcal{O}_3\square^{n-1}\partial^\ell\mathcal{O}_3:$ which have the same conformal twist when $n\geq 1$ and lead to a degeneracy. We will discuss more about operator mixing in Section \ref{Sec:loopAdS5S3}.} We will use this simplified model to explore the analytic structure of super gluon one-loop amplitudes in AdS, and to see what color structures can arise. We will first perform  calculations in the $E_8$ theory where we use the explicit crossing matrices of $E_8$ and apply the AdS unitarity method in each independent flavor channel. The AdS unitarity method was first developed in \cite{Aharony:2016dwx}. Here we will use the Mellin version of this method, originally presented in  \cite{Aharony:2016dwx} and further developed in \cite{Alday:2018kkw,Alday:2019nin}, to manifest the simple analytic structure of the one-loop amplitude. After solving the $E_8$ example, we show the result can be cast in a form agnostic of the color group choice. We then prove that this form of the answer holds for general gauge groups.

\subsection{One-loop amplitude with $E_8$ color group}
As was pointed out in \cite{Aharony:2016dwx}, one-loop correlators are determined by the $\log^2U$ coefficients in the small $U$ expansion. Our first task is to compute these coefficients in each independent color channel.

Our starting point is the disconnected correlator. We use (\ref{P1Padj}) to express $\delta^{I_1I_4}\delta^{I_2I_3}$ and $\delta^{I_1I_3}\delta^{I_2I_4}$  as t- and u-channel projectors, and then the crossing matrices (\ref{ttos}) and (\ref{utos}) to decompose them in the s-channel. Using the results in Section \ref{sec:scfblocks}, we find that the contributions of long operators are the same in channels of the same parity
\begin{eqnarray}
\nonumber &&\mathcal{H}^{(0)}_{2222,{\rm long},{\bf 1}}=\mathcal{H}^{(0)}_{2222,{\rm long},{\bf 3875}}=\mathcal{H}^{(0)}_{2222,{\rm long},{\bf 27000}}=\sum_{n,\ell=0}^\infty \left(\tfrac{1+(-1)^\ell}{2}\right) (8A_{n,\ell}^{(2)})U^{-1}G_{6+2n+\ell,\ell}(z,\bar{z})\;,\\
&&\mathcal{H}^{(0)}_{2222,{\rm long},{\bf 248}}=\mathcal{H}^{(0)}_{2222,{\rm long},{\bf 30380}}=\sum_{n,\ell=0}^\infty \left(\tfrac{1-(-1)^\ell}{2}\right) (8A_{n,\ell}^{(2)})U^{-1}G_{6+2n+\ell,\ell}(z,\bar{z})\;.
\end{eqnarray}
Similarly, at tree level we use (\ref{P1Padj}), (\ref{ctoP}), (\ref{ttos}) and (\ref{utos}) to find the following s-channel decompositions of the $\log U$ coefficients
\begin{equation}
\left(\begin{array}{l}\mathcal{H}^{(1)}_{2222,{\bf 1}}\big|_{\log U} \\\mathcal{H}^{(1)}_{2222,{\bf 3875}}\big|_{\log U} \\\mathcal{H}^{(1)}_{2222,{\bf 27000}}\big|_{\log U}\end{array}\right)=(C_{2,2,2})^2\psi^2h^\vee\left(\begin{array}{c}-3 \\-\frac{3}{5} \\\frac{1}{10}\end{array}\right)\times \sum_{n,\ell=0}^\infty\left(\tfrac{1+(-1)^\ell}{2}\right)\omega_{n,\ell}^{(2)} U^{-1} G_{6+2n+\ell,\ell}(z,\bar{z})\;,
\end{equation}
\begin{equation}
\left(\begin{array}{l}\mathcal{H}^{(1)}_{2222,{\bf 248}}\big|_{\log U} \\\mathcal{H}^{(1)}_{2222,{\bf 30380}}\big|_{\log U}\end{array}\right)=(C_{2,2,2})^2\psi^2h^\vee\left(\begin{array}{c}\frac{3}{2} \\0\end{array}\right)\times \sum_{n,\ell=0}^\infty\left(\tfrac{1-(-1)^\ell}{2}\right)\omega_{n,\ell}^{(2)} U^{-1} G_{6+2n+\ell,\ell}(z,\bar{z})\;.
\end{equation}
Note that in the adjoint (${\bf 248}$) channel, there are two types of contributions. The $\log U$ coefficient receives contributions not only from the two crossed channels, but also from the s-channel. 

Because there is no operator mixing, the $\log^2U$ coefficients at one loop are simply given by squaring the tree-level coefficients and then dividing by the disconnected coefficients. We find that they can be written as
\begin{equation}
\left(\begin{array}{l}\mathcal{H}^{(2),{\rm 5d}}_{2222,{\bf 1}}\big|_{\log^2 U} \\\mathcal{H}^{(2),{\rm 5d}}_{2222,{\bf 3875}}\big|_{\log^2 U} \\\mathcal{H}^{(2),{\rm 5d}}_{2222,{\bf 27000}}\big|_{\log^2 U}\end{array}\right)=(C_{2,2,2})^4(\psi^2h^\vee)^2\left(\begin{array}{c}9 \\\frac{9}{25} \\\frac{1}{100}\end{array}\right)\times \mathcal{F}_{\rm even}(z,\bar{z})\;,
\end{equation}
\begin{equation}
\left(\begin{array}{l}\mathcal{H}^{(2),{\rm 5d}}_{2222,{\bf 248}}\big|_{\log^2 U} \\\mathcal{H}^{(2),{\rm 5d}}_{2222,{\bf 30380}}\big|_{\log^2 U}\end{array}\right)=(C_{2,2,2})^4(\psi^2h^\vee)^2\left(\begin{array}{c}\frac{9}{4} \\ 0 \end{array}\right)\times \mathcal{F}_{\rm odd}(z,\bar{z})\;,
\end{equation}
where the two basic functions of cross ratios are
\begin{equation}\label{defcalF}
\begin{split}
 \mathcal{F}_{\rm even}(z,\bar{z})=&\sum_{n,\ell=0}^\infty \frac{1}{2}\frac{(\omega_{n,\ell}^{(2)})^2}{8A^{(2)}_{n,\ell}}\left(\tfrac{1+(-1)^\ell}{2}\right)U^{-1}G_{6+2n+\ell,\ell}(z,\bar{z})\;,\\
 \mathcal{F}_{\rm odd}(z,\bar{z})=&\sum_{n,\ell=0}^\infty \frac{1}{2}\frac{(\omega_{n,\ell}^{(2)})^2}{8A^{(2)}_{n,\ell}}\left(\tfrac{1-(-1)^\ell}{2}\right)U^{-1}G_{6+2n+\ell,\ell}(z,\bar{z})\;.
 \end{split}
\end{equation}

We now follow the strategy of \cite{Alday:2018kkw,Alday:2019nin} and look for a Mellin amplitude that reproduces these $\log^2U$ coefficients. To achieve this, we look at the small $V$ expansion of $\mathcal{F}_{\rm even}$ and $\mathcal{F}_{\rm odd}$. We find that they contain at most $\log^2V$ singularities in the $V\to 0$ limit. Such  $\log^2U\log^2V$ singularities of the four-point correlator require the reduced Mellin amplitudes to have simultaneous poles of the form\footnote{Recall that the Gamma function factor already have double poles at these locations. The inverse Mellin integrand then has in total triple poles, which give rise to $\log^2U\log^2V$ singularities upon taking residues.}
\begin{equation}
\frac{1}{(s-2m)(t-2n)}\;,\quad m,n=2,3,4,\ldots\;.
\end{equation}
 We then make the following simple ansatz which is consistent with the symmetry in the even and odd channels
\begin{eqnarray}\label{Mtildeevenodd5d}
\nonumber \widetilde{\mathcal{M}}_{\rm even}&=&\sum_{m,n=2}^\infty c^{\rm even}_{mn}\left(\frac{1}{(s-2m)(t-2n)}+\frac{1}{(s-2m)(\tilde{u}-2n)}\right)\;,\\
\widetilde{\mathcal{M}}_{\rm odd}&=&\sum_{m,n=2}^\infty c^{\rm odd}_{mn}\left(\frac{1}{(s-2m)(t-2n)}-\frac{1}{(s-2m)(\tilde{u}-2n)}\right)\;.
\end{eqnarray}
Here we have assumed that $c^{\rm even}_{mn}$ and $c^{\rm odd}_{mn}$ are numbers independent of Mandelstam variables. Taking the residues in the inverse Mellin transformation, we can extract the $\log^2U\log^2V$ coefficients as a power expansion in $U$ and $V$. By matching them with the $\log^2V$ coefficients $\mathcal{F}_{\rm even, odd}$ in the small $U$, $V$ expansion, we can obtain the coefficients $c^{\rm even,odd}_{mn}$. We find they are the same function up to a sign
\begin{equation}
c^{\rm odd}_{mn}=-c^{\rm even}_{mn}=-c^{\rm 5d}_{mn}\;,
\end{equation}
and
\begin{equation}\label{cmnAdS5}
\begin{split}
c_{mn}^{\rm 5d}=&\frac{\sqrt{\pi } \Gamma (m-1) \Gamma (n-1) \Gamma \left(m+n-\frac{5}{2}\right)}{18 \Gamma \left(m+\frac{1}{2}\right) \Gamma \left(n+\frac{1}{2}\right) \Gamma (m+n-1)}\\
&\times \left(3 m^2 n-2 m^2+3 m n^2-11 m n+6 m-2 n^2+6 n-3\right)
\end{split}
\end{equation}
is symmetric under $n\leftrightarrow m$. Note that so far we have only used and matched the $\log^2V$ coefficients of $\mathcal{F}_{\rm even, odd}$. However, it turns out that the amplitudes (\ref{Mtildeevenodd5d}) with these coefficients reproduce the {\it full} functions $\mathcal{F}_{\rm even, odd}$, which include $\log V$ singularities and regular terms as well. This implies that there are no additional single poles in $s$ in the reduced Mellin amplitude, confirming the assumption that $c^{\rm even}_{mn}$ and $c^{\rm odd}_{mn}$ are independent of the Mandelstam variables. 

Let us denote the above building block amplitudes as 
\begin{eqnarray}\label{B5d}
\nonumber \mathcal{B}^{\rm 5d}_{st}=\sum_{m,n=2}^\infty\frac{c_{mn}^{\rm 5d}}{(s-2m)(t-2n)}\;,\\
\mathcal{B}^{\rm 5d}_{su}=\sum_{m,n=2}^\infty\frac{c_{mn}^{\rm 5d}}{(s-2m)(\tilde{u}-2n)}\;,\\
\nonumber \mathcal{B}^{\rm 5d}_{tu}=\sum_{m,n=2}^\infty\frac{c_{mn}^{\rm 5d}}{(t-2m)(\tilde{u}-2n)}\;,
\end{eqnarray}
which are the $AdS_5$ {\it box diagrams}. This interpretation of the amplitudes becomes clear when we examine their behavior in the flat space limit. The flat space limit corresponds to the high energy regime of the Mellin amplitude where $s$, $t$ are taken to be large \cite{Penedones:2010ue}. As we show in Appendix \ref{App:flatspacebox}, any Mellin amplitude which has the form of $\nonumber \mathcal{B}^{\rm 5d}_{st}$ and coefficients with the following scaling behavior
\begin{equation}
c^{\rm 5d}_{mn} = \frac{(m n)^{\frac{D}{2}-3}}{(m+n)^{\frac{D}{2}-2}} + \cdots
\end{equation}
in the large $m$, $n$ limit, becomes the $D$-dimensional scalar one-loop box diagram in flat space. In the current case the coefficients (\ref{cmnAdS5}) scale as $1/(mn(m+n))^{\frac{1}{2}}$ and therefore (\ref{B5d}) become the 5d box diagrams in the flat space limit.

Continuing the above discussion, the one-loop reduced amplitude of $\langle 2222\rangle$ should be
\begin{equation}\label{oneloopE8}
\widetilde{\mathcal{M}}_{2222}^{AdS_5}=(C_{2,2,2})^4(\psi^2h^\vee)^2\left(\begin{array}{c}9( \mathcal{B}^{\rm 5d}_{st}+\mathcal{B}^{\rm 5d}_{su}+\lambda_1 \mathcal{B}^{\rm 5d}_{tu}) \\\frac{9}{25}( \mathcal{B}^{\rm 5d}_{st}+\mathcal{B}^{\rm 5d}_{su}+\lambda_2 \mathcal{B}^{\rm 5d}_{tu}) \\\frac{1}{100}( \mathcal{B}^{\rm 5d}_{st}+\mathcal{B}^{\rm 5d}_{su}+\lambda_3 \mathcal{B}^{\rm 5d}_{tu}) \\\frac{9}{4}( \mathcal{B}^{\rm 5d}_{su}-\mathcal{B}^{\rm 5d}_{st}) \\0\end{array}\right)\;,
\end{equation}
in order to be consistent with the $\log^2U$ coefficients. Note that in the symmetric channels, we have also included the box diagram $\mathcal{B}^{\rm 5d}_{tu}$ which is compatible with the symmetry of these channels. This box diagram does not contribute the $\log^2U$ coefficients, and therefore the parameters $\lambda_i$ are not fixed by them. However, the one-loop reduced amplitude should also be crossing symmetric
\begin{eqnarray}
\mathcal{M}_{2222,a}^{AdS_5}(s,t)&=&\sum_{a'} ({\rm F}_t)_a{}^{a'}\mathcal{M}_{2222,a'}^{AdS_5}(t,s)\;,\\
\mathcal{M}_{2222,a}^{AdS_5}(s,t)&=&\sum_{a'} ({\rm F}_u)_a{}^{a'}\mathcal{M}_{2222,a'}^{AdS_5}(u,t)\;.
\end{eqnarray}
Imposing these conditions uniquely solves the unknown coefficients, and gives 
\begin{equation}
\lambda_1=\frac{1}{2}\;,\quad \lambda_2=-\frac{3}{2}\;,\quad \lambda_3=16\;.
\end{equation}
Let us also make a comment regarding the completeness of the ansatz. As pointed out earlier, the fact that the reduced Mellin amplitude reproduces the complete $\log^2U$ coefficient rules out single poles in the $s$ variable. Via the full crossing symmetry, this also rules out single poles in $t$ and $\tilde{u}$. Therefore, the only ambiguities are regular contact terms.

\begin{figure}[h]
\centering
\includegraphics[width=0.95\textwidth]{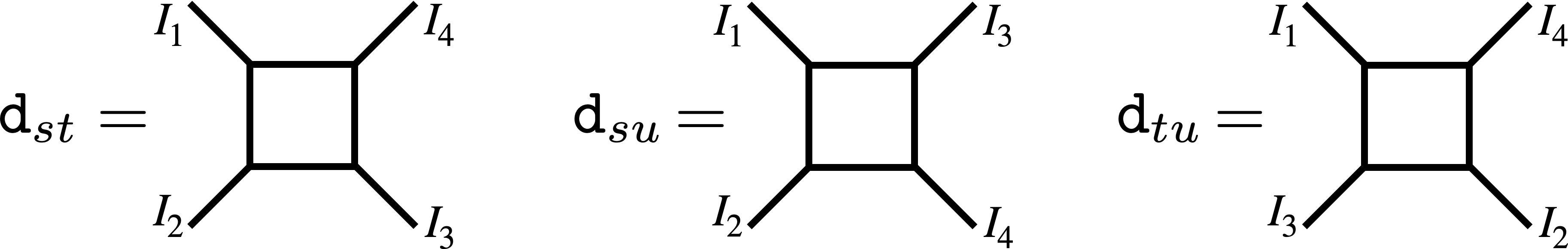}
\caption{Box diagrams for colors. Each vertex represents a structure constant and the lines are in the adjoint representation.}
    \label{fig:colorbox}
\end{figure}

Written as (\ref{oneloopE8}), the color structure of the one-loop amplitude is quite obscure and the answer appears to sensitively depend on the specific choice of the gauge group. However, we claim that the amplitude can be rewritten in a more illuminating form as  
\begin{equation}\label{oneloopE8box}
\widetilde{\mathcal{M}}_{2222}^{AdS_5}=9(C_{2,2,2})^4\big(\mathtt{d}_{st}\mathcal{B}_{st}^{\rm  5d}+\mathtt{d}_{su}\mathcal{B}_{su}^{\rm  5d}+\mathtt{d}_{tu}\mathcal{B}_{tu}^{\rm  5d}\big)
\end{equation}
where 
\begin{equation}
\begin{split}
\mathtt{d}_{st}=& f^{JI_1K}f^{KI_2L}f^{LI_3M}f^{MI_4J}\;,\\
\mathtt{d}_{su}=& f^{JI_1K}f^{KI_2L}f^{LI_4M}f^{MI_3J}\;,\\
\mathtt{d}_{tu}=& f^{JI_1K}f^{KI_3L}f^{LI_2M}f^{MI_4J}\;,
\end{split}
\end{equation}
are box diagrams for the color part (see Figure \ref{fig:colorbox}). Clearly, this form of the answer can be generalized to any color group. 

\begin{figure}[h]
\centering
\includegraphics[width=0.85\textwidth]{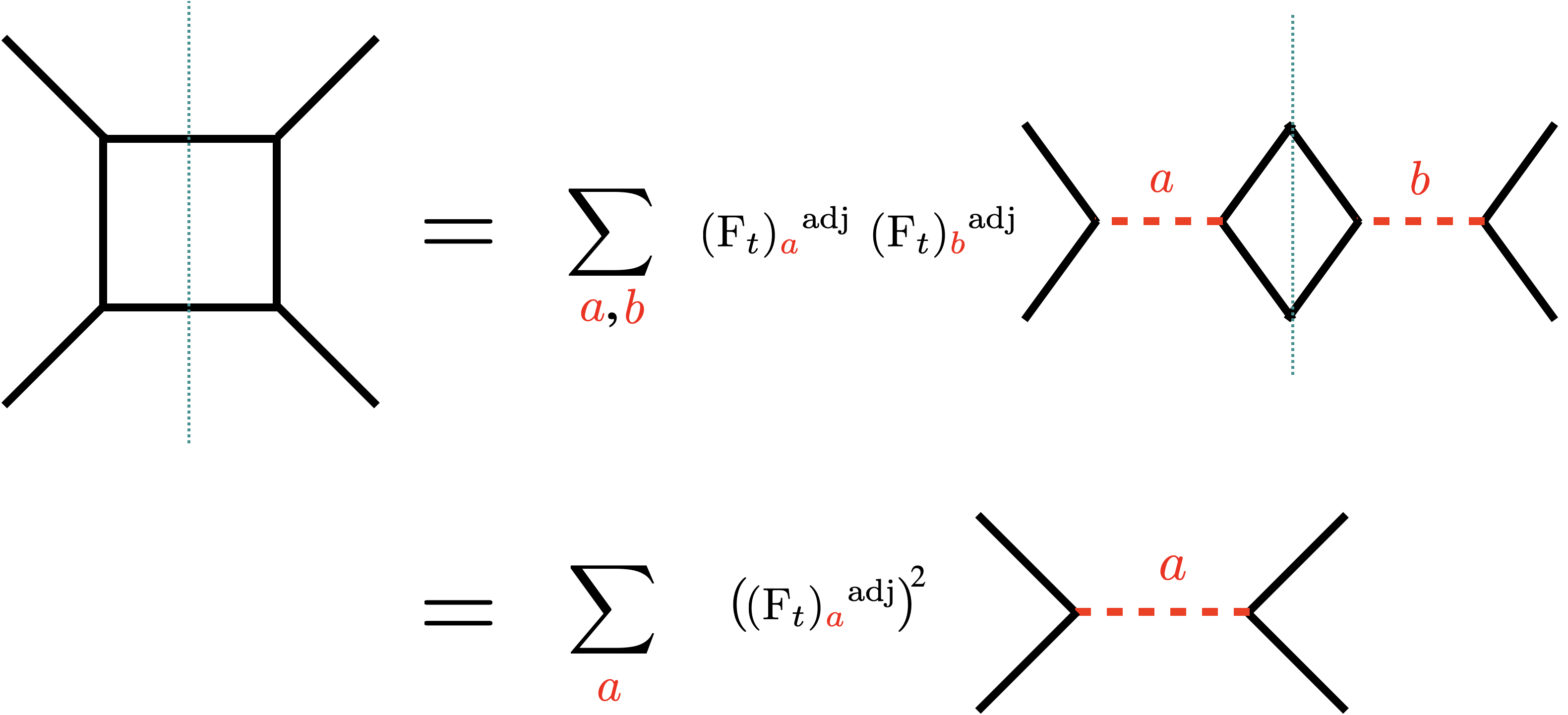}
\caption{Writing the color box function $\mathcal{B}_{st}$ as s-channel projectors. The box diagram can be obtained as the product of two t-channel projectors in which the adjoint representations are exchanged. We then use the crossing matrix to expand each t-channel projector in the s-channel, and idempotence of projectors to write the function as a linear combination of s-channel projectors.}
    \label{fig:boxtotree}
\end{figure}

To show that the two representations (\ref{oneloopE8}) and (\ref{oneloopE8box}) are equivalent, we decompose the color structures $\mathtt{d}_{st}$, $\mathtt{d}_{st}$, $\mathtt{d}_{tu}$ into s-channel projectors and compare the coefficients in each channel. This can be done by following the procedure outlined in Figure \ref{fig:boxtotree}. We first cut open the box diagram vertically, and view it as the contraction of two t-channel projectors associated with the adjoint representation. We then use the crossing matrix to express each t-channel projector as a linear combination of the s-channel projectors. Finally, using the idempotence of projectors (\ref{idempo}) we reduce the color box to be a linear combination of s-channel projectors, and find
\begin{equation}\label{BstP}
\mathtt{d}_{st}=(\psi^2h^\vee)^2\sum_a \big(({\rm F}_t)_a{}^{\rm adj}\big)^2 {\rm P}^{I_1I_2|I_3I_4}_a\;.
\end{equation}
The other two color boxes can be obtained by substituting $I_3\leftrightarrow I_4$ and $I_2\leftrightarrow I_3$ respectively and then applying crossing relations of projectors. Using these relations and the explicit form of the $E_8$ crossing matrices, it is straightforward to verify that (\ref{oneloopE8}) and (\ref{oneloopE8box}) agree.

\subsection{One-loop amplitude with general color group}\label{sec:oneloopgeneralGF}
In the previous subsection we worked out the explicit example of $E_8$ gauge group. The answer was then rewritten in a different form  (\ref{oneloopE8box}), which depends only on the structure constants of the group and indicates a straightforward generalization to any color group. In this subsection, we prove that the one-loop reduced amplitude for general color groups is indeed given by (\ref{oneloopE8box}).

Let us first start with some simple facts about projectors. From (\ref{idempo}), (\ref{Pdimform}) and the definition (\ref{P1Padj}), it follows that 
\begin{equation}
({\rm F}_t)_a{}^{\bf 1}=\frac{1}{{\rm dim} (G_F)}
\end{equation}
for any representation $a$ in the tensor product. Similarly, we also have
\begin{equation}
({\rm F}_u)_a{}^{\bf 1}=\frac{(-1)^{{\rm R}_a}}{{\rm dim} (G_F)}\;.
\end{equation}
Therefore, the contribution of the disconnected correlator is always 
\begin{equation}
\mathcal{H}^{(0)}_{2222,{\rm long},{\rm even}}=\sum_{n,\ell=0}^\infty \left(\tfrac{1+(-1)^\ell}{2}\right) (8A_{n,\ell}^{(2)})U^{-1}G_{6+2n+\ell,\ell}(z,\bar{z})
\end{equation}
for the even channels, and  
\begin{equation}
\mathcal{H}^{(0)}_{2222,{\rm long},{\rm odd}}=\sum_{n,\ell=0}^\infty \left(\tfrac{1-(-1)^\ell}{2}\right) (8A_{n,\ell}^{(2)})U^{-1}G_{6+2n+\ell,\ell}(z,\bar{z})
\end{equation}
for the odd channels, regardless of the gauge group $G_F$. Note that the $1/{\rm dim}(G_F)$ factor is cancelled out due to the appearance of the same factor in the definition (\ref{P1Padj}).

\begin{figure}[h]
\centering
\includegraphics[width=0.7\textwidth]{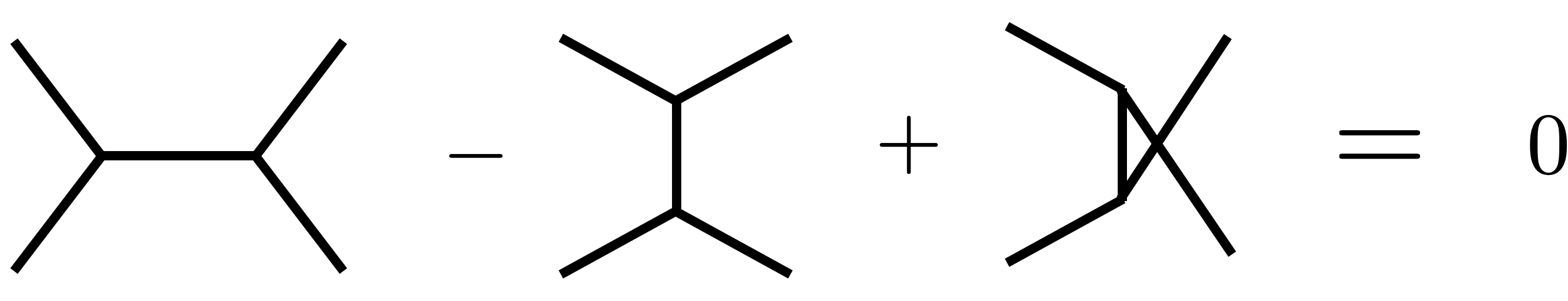}
\caption{Jacobi identity in the diagrammatic form. The second term has a  minus sign because vertices are oriented.}
    \label{fig:jacobi}
\end{figure}

At tree level, the t- and u-channel exchanges $\mathcal{H}_{22kk}^{(t)}\big|_{\log U}$ and $\mathcal{H}_{22kk}^{(u)}\big|_{\log U}$ together contribute
\begin{equation}\label{AdS5treeeven}
(C_{2,2,2})^2\psi^2h^\vee \times (-3 ({\rm F}_t)_a{}^{\rm adj}) \sum_{n,\ell=0}^\infty\left(\tfrac{1+(-1)^\ell}{2}\right)\omega_{n,\ell}^{(2)} U^{-1} G_{6+2n+\ell,\ell}(z,\bar{z})\;,
\end{equation}
in the even channels, and
\begin{equation}
(C_{2,2,2})^2\psi^2h^\vee \times ({\rm F}_t)_a{}^{\rm adj} \sum_{n,\ell=0}^\infty\left(\tfrac{1-(-1)^\ell}{2}\right)\omega_{n,\ell}^{(2)} U^{-1} G_{6+2n+\ell,\ell}(z,\bar{z})\;,
\end{equation}
in the odd channels. To proceed, we need to compute the $({\rm F}_t)_a{}^{\rm adj}$ matrix elements for $a$ odd. We look at the Jacobi identity which is diagrammatically represented by Figure \ref{fig:jacobi}. We contract both sides with the s-channel projector ${\rm P}^{I_1I_2|I_3I_4}_a$. The second and third terms respectively give $-({\rm F}_t)_a{}^{\rm adj}$ and $-({\rm F}_u)_a{}^{\rm adj}$. Because $a$ is odd, the last two terms contribute equally. On the other hand, due to the delta function in (\ref{idempo}) the first term is only nonzero and equals to one when $a$ is the adjoint representation. We then find
\begin{equation}
({\rm F}_t)_{\rm adj}{}^{\rm adj}=({\rm F}_u)_{\rm adj}{}^{\rm adj}=\frac{1}{2}\;,
\end{equation}
and 
\begin{equation}
({\rm F}_t)_{\rm adj}{}^{\rm a}=({\rm F}_u)_{\rm adj}{}^{\rm a}=0\;,\quad a\in {\rm odd}\;,\;\; a\neq {\rm adj}\;.
\end{equation}
On the other hand, we note that in the adjoint channel there is also another contribution from the s-channel tree-level exchange $\mathcal{H}_{22kk}^{(s)}\big|_{\log U}$. This allows us to write the total contribution in the odd channels as
\begin{equation}
(C_{2,2,2})^2\psi^2h^\vee \times (3({\rm F}_t)_a{}^{\rm adj}) \sum_{n,\ell=0}^\infty\left(\tfrac{1-(-1)^\ell}{2}\right)\omega_{n,\ell}^{(2)} U^{-1}G_{6+2n+\ell,\ell}(z,\bar{z})\;,
\end{equation}
which have the same form as the even channels (\ref{AdS5treeeven}) up to an overall minus sign.

Let us now move on to the one-loop level. It follows from the above discussion that the $\log^2 U$ coefficients are
\begin{equation}\label{AdS5log2U}
\mathcal{H}^{(2),{\rm 5d}}_{2222,a}\big|_{\log^2 U}=9(C_{2,2,2})^4(\psi^2h^\vee)^2(({\rm F}_t)_a{}^{\rm adj})^2\times \left\{\begin{array}{c}\mathcal{F}_{\rm even}\,,\;\;a\in{\rm even} \\ \mathcal{F}_{\rm odd}\,,\;\;a\in{\rm odd}\end{array}\right.
\end{equation}
where $\mathcal{F}_{\rm even, odd}$ were defined in (\ref{defcalF}). 
Then from (\ref{BstP}) and the discussion of the Mellin amplitudes of $\mathcal{F}_{\rm even, odd}$, we know that $\widetilde{\mathcal{M}}_{2222}^{AdS_5}$ must include $9(C_{2,2,2})^4(\mathtt{d}_{st}\mathcal{B}_{st}^{\rm  5d}+\mathtt{d}_{su}\mathcal{B}_{su}^{\rm  5d})$ as part of the answer. By requiring that the one-loop reduced amplitude is crossing symmetric, we arrive at (\ref{oneloopE8box}). Note that in the above derivation, we have not used any special properties that are specific to certain groups. Therefore, (\ref{oneloopE8box}) is the one-loop reduced Mellin amplitude for any gauge group.

\section{One-loop gluon amplitude on  AdS$_5$$\times$S$^3$}\label{Sec:loopAdS5S3}
Let us now compute the super gluon one-loop amplitudes in the theories of D3-branes probing F-theory singularities. Compared with the $AdS_5$ toy model studied in the previous section, here the space is $AdS_5\times S^3$. The $S^3$ factor gives rise to infinitely many additional massive super gluons from Kaluza-Klein reduction, which must be summed over in the one-loop amplitude. The existence of these massive modes leads to a degeneracy in double-trace operators, and presents additional complexity in using the AdS unitarity method. We will first review how to solve this mixing problem, and then compute the one-loop reduced Mellin amplitude for an arbitrary gauge group $G_F$. A remarkable feature of our result is that it admits a closed form expression, unlike the case without $S^3$. We will also show that the answer reproduces the 8d box diagrams in the flat space limit, which agrees with the fact that our background is now eight dimensional. 

\subsection{Mixing problem}
For a fixed engineering conformal twist $\tau=4+2n$ and Lorentz spin $\ell$, the space of $SU(2)_R\times SU(2)_L$ invariant double-trace operators in each flavor channel is $(n+1)$-dimensional. Since operators with different flavor representations do not mix, we will suppress the flavor indices   in this subsection to simplify the notation. Schematically, these operators can be written as 
\begin{equation}
:\mathcal{O}_2\square^n\partial^\ell \mathcal{O}_2:\;,\;\; :\mathcal{O}_3\square^{n-1}\partial^\ell \mathcal{O}_3:\;,\;\; \ldots\;\; :\mathcal{O}_{n+2}\partial^\ell \mathcal{O}_{n+2}:
\end{equation}
where we have also suppressed the $SU(2)_R\times SU(2)_L$ indices. For computing the $\langle2222\rangle$ one-loop amplitude, only the singlets are relevant. Turning on interactions lifts the degeneracies, and we will denote the eigenstates $\mathbb{O}_{n,\ell,i}$.

At the disconnected level, these double-trace operators contribute to the reduced correlator $\langle kkkk\rangle$ as
\begin{equation}
\mathcal{H}^{(0)}_{kkkk,{\rm long}}=\sum_{n=0}^\infty\sum_{\ell} a_{n,\ell}^{(k)}U^{-1}G_{2k+2n+2+\ell,\ell}(z,\bar{z})
\end{equation}
where $a_{n,\ell}^{(k)}$ are related to three-point function coefficients by
\begin{equation}
a_{n,\ell}^{(k)}=\sum_{i=1}^{k+n-1}C^{(0)}_{kk\mathbb{O}_{k+n-2,\ell,i}}C^{(0)}_{kk\mathbb{O}_{k+n-2,\ell,i}}\;.
\end{equation}
We remind the reader that there is a shift in dimension by 2 in the conformal blocks according to (\ref{longK}) because we are looking at the reduced correlators. At tree level, we will focus on the $\log U$ coefficients of the reduced correlator of $\langle 22kk \rangle$. They can be decomposed into long operator conformal blocks as
\begin{equation}
\mathcal{H}^{(1)}_{22kk}\big|_{\log U}=\sum_{n=0}^\infty\sum_{\ell} \omega^{(k)}_{n,\ell} U^{-1} G_{2k+2n+2+\ell,\ell}(z,\bar{z})
\end{equation}
where $\omega^{(k)}_{n,\ell}$ is given by 
\begin{equation}
\omega^{(k)}_{n,\ell}=\sum_{i=1}^{k+n-1}C^{(0)}_{22\mathbb{O}_{k+n-2,\ell,i}}\gamma^{(1)}_{k+n-2,\ell,i}C^{(0)}_{kk\mathbb{O}_{k+n-2,\ell,i}}
\end{equation}
with $\gamma^{(1)}_{k+n-2,\ell,i}$ being the tree-level anomalous dimension of $\mathbb{O}_{k+n-2,\ell,i}$.

Let us now look at the $\langle2222\rangle$ correlator at  one loop.  The key object is the $\log^2U$ coefficient of its reduced correlator, which is computed by
\begin{equation}
\mathcal{H}^{(2)}_{2222}\big|_{\log^2 U}=\sum_{n=0}^\infty\sum_{\ell} F_{n,\ell} U^{-1} G_{6+2n+\ell,\ell}(z,\bar{z})\;,
\end{equation}
with
\begin{equation}\label{halfagamma2}
F_{n,\ell} =\frac{1}{2}\sum_{i=1}^{n+1}C^{(0)}_{22\mathbb{O}_{n,\ell,i}}\big(\gamma^{(1)}_{n,\ell,i}\big)^2C^{(0)}_{22\mathbb{O}_{n,\ell,i}}\;.
\end{equation}
Due to operator mixing, we cannot compute $F_{n,\ell}$ using only the $\langle 2222\rangle$ correlator at disconnected and tree levels. However, this quantity can still be expressed in terms of the lower-order data extracted from $\mathcal{H}^{(0)}_{kkkk,{\rm long}}$ and $\mathcal{H}^{(1)}_{22kk}\big|_{\log U}$ if we consider arbitrary $k$. To see this, let us rewrite the above results in matrix notation. We first define a $(q+1)$ by $(q+1)$ matrix
\begin{equation}
\mathbf{C}_{q,\ell}=\left(\begin{array}{c} C^{(0)}_{22\mathbb{O}_{q,\ell,i}} \\\ldots \\C^{(0)}_{q+2,q+2\mathbb{O}_{q,\ell,i}}\end{array}\right)\;,
\end{equation}
which involves the double-trace operators with conformal twist $2q+4$. Then $\mathbf{C}\mathbf{C}^T$ gives the long multiplet coefficients of the following matrix of disconnected correlators
\begin{equation}
\mathbf{C}_{q,\ell}\big(\mathbf{C}_{q,\ell}\big)^T=\left(\begin{array}{ccc}\langle2222\rangle^{(0)} & \ldots & \langle2,2,q+2,q+2\rangle^{(0)} \\\ldots & \ldots & \ldots \\\langle q+2,q+2,2,2\rangle^{(0)} & \ldots & \langle q+2,q+2,q+2,q+2\rangle^{(0)}\end{array}\right)_{\text{long coef.}}
\end{equation}
Note that due to large $N$ factorization, $\langle rrss\rangle^{(0)}$ contains no long operators unless $r=s$. Therefore, the matrix $\mathbf{C}\mathbf{C}^T$ is {\it diagonal}, and we will denote it as 
\begin{equation}
\mathbf{C}_{q,\ell}\big(\mathbf{C}_{q,\ell}\big)^T=\mathbf{N}_{q,\ell}\;.
\end{equation}
Moreover, we can collect the $q+1$ eigenvalues of tree-level anomalous dimensions into a diagonal matrix
\begin{equation}
\mathbf{\Gamma}_{q,\ell}={\rm diag}(\gamma^{(1)}_{q,\ell,1},\ldots \gamma^{(1)}_{q,\ell,q+1})\;.
\end{equation}
We can then consider the matrix
\begin{equation}
\mathbf{C}_{q,\ell} \mathbf{\Gamma}_{q,\ell} \big(\mathbf{C}_{q,\ell}\big)^T=\left(\begin{array}{ccc}\langle2222\rangle^{(1)} & \ldots & \langle2,2,q+2,q+2\rangle^{(1)} \\\ldots & \ldots & \ldots \\\langle q+2,q+2,2,2\rangle^{(1)} & \ldots & \langle q+2,q+2,q+2,q+2\rangle^{(1)}\end{array}\right)_{\log U\text{ coef.}}
\end{equation}
The first row (or the first column) are $\omega^{(k)}_{q-k+2,\ell}$ with $k=2,3,\ldots q+2$, which appear in the tree-level $\langle 22kk\rangle$ correlators. Furthermore, it is easy to see that quantity (\ref{halfagamma2}) corresponds to the matrix elements 
\begin{equation}
F_{n,\ell}=\frac{1}{2}\bigg(\mathbf{C}_{n,\ell} \mathbf{\Gamma}_{n,\ell}  \big(\mathbf{C}_{n,\ell}\big)^T\bigg) \mathbf{N}_{n,\ell}^{-1} \bigg(\mathbf{C}_{n,\ell} \mathbf{\Gamma}_{n,\ell}  \big(\mathbf{C}_{n,\ell}\big)^T\bigg)\;.
\end{equation}
Since $\mathbf{N}$ is diagonal, it follows that this matrix element can be expressed as 
\begin{equation}\label{Fnl}
F_{n,\ell}=\frac{1}{2}\sum_{k=2}^{n+2}\frac{\big(\omega^{(k)}_{n-k+2,\ell}\big)^2}{a^{(k)}_{n-k+2,\ell}}\;.
\end{equation}
This gives the correct answer for the average of the squared tree-level anomalous dimensions, with operator mixing properly taken into account.
 
\subsection{One-loop amplitude}\label{subsecAdS5S3oneloop}
Because the color structure is independent of mixing, the discussion in Section \ref{sec:oneloopgeneralGF} directly applies here.  The $\log^2U$ coefficients of $AdS_5\times S^3$ four-point correlator take the same form as in the $AdS_5$ case (\ref{AdS5log2U})
\begin{equation}\label{AdS5S3log2U}
\mathcal{H}^{(2),{\rm 8d}}_{2222,a}\big|_{\log^2 U}=9(C_{2,2,2})^2(\psi^2h^\vee)^2(({\rm F}_t)_a{}^{\rm adj})^2\times \left\{\begin{array}{c}\mathcal{F}_{\rm even}^{\rm 8d}\,,\;\;a\in{\rm even} \\ \mathcal{F}_{\rm odd}^{\rm 8d}\,,\;\;a\in{\rm odd}\end{array}\right.\;.
\end{equation}
The only difference is that $\mathcal{F}_{\rm even,odd}$ are now replaced by 
\begin{equation}\label{defcalF8d}
\begin{split}
 \mathcal{F}^{\rm 8d}_{\rm even}(z,\bar{z})=&\sum_{n,\ell=0}^\infty F_{n,\ell} \left(\tfrac{1+(-1)^\ell}{2}\right) U^{-1} G_{6+2n+\ell,\ell}(z,\bar{z})\;,\\
 \mathcal{F}^{\rm 8d}_{\rm odd}(z,\bar{z})=&\sum_{n,\ell=0}^\infty F_{n,\ell} \left(\tfrac{1-(-1)^\ell}{2}\right) U^{-1} G_{6+2n+\ell,\ell}(z,\bar{z})\;.
 \end{split}
\end{equation}
where $F_{n,\ell}$ is given by (\ref{Fnl}) as
\begin{equation}
F_{n,\ell}=\sum_{k=2}^{n+2} E_{k,n,\ell}\;,
\end{equation}
with 
\begin{equation}
\begin{split}
E_{k,n,\ell}=\frac{1}{2}\times\frac{\left(\omega^{(k)}_{n+2-k,\ell}\right)^2}{8A^{(k)}_{n+2-k,\ell}}=&\frac{\pi  (k-1)^2 (n+1) 2^{-4 n-\ell -6} (n+\ell +2)  }{9 (\ell +1) (2 n+\ell +4) \Gamma \left(n+\frac{3}{2}\right) }\\
\times &\frac{\Gamma (n+3) \Gamma (k+n+1) \Gamma (n+\ell +4) \Gamma (-k+n+\ell +4)}{\Gamma (-k+n+3) \Gamma \left(n+\ell +\frac{5}{2}\right) \Gamma (k+n+\ell +2)}\;.
\end{split}
\end{equation}
Following the same procedure, we find that the $\log^2V$ coefficients of $\mathcal{F}^{\rm 8d}_{\rm even}$ and $\mathcal{F}^{\rm 8d}_{\rm odd}$ can be respectively matched by the reduced Mellin amplitudes
\begin{equation}\label{Mt8deo}
\widetilde{\mathcal{M}}^{\rm 8d}_{\rm even}=\mathcal{B}^{\rm 8d}_{st}+\mathcal{B}^{\rm 8d}_{su}\;,\quad\quad \widetilde{\mathcal{M}}^{\rm 8d}_{\rm odd}=\mathcal{B}^{\rm 8d}_{su}-\mathcal{B}^{\rm 8d}_{st}\;.
\end{equation}
Here we have defined the eight dimensional $AdS^5\times S^3$ box diagrams as 
\begin{eqnarray}\label{boxAdS5S3}
\nonumber \mathcal{B}^{\rm 8d}_{st}=\sum_{m,n=2}^\infty\frac{c_{mn}^{\rm 8d}}{(s-2m)(t-2n)}\;,\\
\mathcal{B}^{\rm 8d}_{su}=\sum_{m,n=2}^\infty\frac{c_{mn}^{\rm 8d}}{(s-2m)(\tilde{u}-2n)}\;,\\
\nonumber \mathcal{B}^{\rm 8d}_{tu}=\sum_{m,n=2}^\infty\frac{c_{mn}^{\rm 8d}}{(t-2m)(\tilde{u}-2n)}\;,
\end{eqnarray}
with the coefficients given by 
\begin{equation}
c_{mn}^{\rm 8d}=\frac{4 \left(3 m^2 n-4 m^2+3 m n^2-16 m n+15 m-4 n^2+15 n-12\right)}{27 (m+n-4) (m+n-3) (m+n-2)}\;.
\end{equation}
Using this solution we can check that the functions $\mathcal{F}^{\rm 8d}_{\rm even}$ and $\mathcal{F}^{\rm 8d}_{\rm odd}$ are completely reproduced by the amplitudes (\ref{Mt8deo}). This implies that the simultaneous poles are again sufficient. It follows from the discussions in Section \ref{sec:oneloopgeneralGF}  that the one-loop reduced Mellin amplitudes for $AdS_5\times S^3$ are
\begin{equation}\label{oneloopAdS5S3}
\widetilde{\mathcal{M}}_{2222}^{AdS_5\times S^3}=9(C_{2,2,2})^4\big(\mathtt{d}_{st}\mathcal{B}_{st}^{\rm  8d}+\mathtt{d}_{su}\mathcal{B}_{su}^{\rm  8d}+\mathtt{d}_{tu}\mathcal{B}_{tu}^{\rm  8d}\big)\;.
\end{equation}
We also note that the coefficients $c_{mn}^{\rm 8d}$ scale as 
\begin{equation}
c_{mn}^{\rm 8d}\sim \frac{mn}{(m+n)^2}+\ldots
\end{equation}
in the large $m$, $n$ limit. According to the results in Appendix \ref{App:flatspacebox}, the Mellin amplitudes (\ref{boxAdS5S3}) become eight-dimensional box integrals in the flat space limit. There is actually a more explicit way to see this. It turns out that the $AdS_5\times S^3$ box diagrams can be computed in a closed form. The basic building block is the following double sum
\begin{equation}
\Theta(s,t) = \sum_{m,n=1} \frac{1}{(m+n)(s-2m)(t-2n)}
\end{equation}
which can be performed in terms of polygamma functions
\begin{eqnarray}
\Theta(s,t) =&&\frac{1}{4 (s+t)} \bigg\{ \psi ^{(1)}\left(-\frac{s}{2}\right)+ \psi ^{(1)}\left(-\frac{t}{2}\right) -\left[\psi ^{(0)}\left(-\frac{s}{2}\right)\right]^2 \nonumber \\ 
&&\quad\quad\quad\quad\quad-\left[\psi ^{(0)}\left(-\frac{t}{2}\right)\right]^2 + 2\psi ^{(0)}\left(-\frac{s}{2}\right) \psi ^{(0)}\left(-\frac{t}{2}\right) \bigg\} \nonumber \\
&& + \frac{2}{s t (s+t)}-\frac{\pi ^2}{4 (s+t)} -\frac{\psi ^{(0)}\left(-\frac{s}{2}\right)+\gamma}{t (s+t)}-\frac{\psi ^{(0)}\left(-\frac{t}{2}\right)+\gamma}{s (s+t)}\;.
\end{eqnarray}
Here $\gamma$ is the Euler-Mascheroni constant. Mellin amplitudes in various examples, including the ones in this paper, can be written in terms of this building block, upon shifting the $s$, $t$ variables, plus edge terms involving a single sum. For the case at hand the double sum is divergent, but it can be regularized in different ways. The difference between different regularizations is simply a constant. This can be seen as follows. Both $\partial_s \mathcal{B}^{\rm 8d}_{st}$ and $\partial_t \mathcal{B}^{\rm 8d}_{st}$ are given by convergent sums, and they lead to $\mathcal{B}^{\rm 8d}_{st}$ up to a constant of integration which depends on a single variable. This together with symmetry under $s \leftrightarrow t$ implies the constant of integration is actually independent of both $s$ and $t$. The final result is given by
\begin{eqnarray}
\mathcal{B}^{\rm 8d}_{st} =& R_0(s,t) \left( \psi ^{(1)}\left(2-\frac{s}{2}\right)+ \psi ^{(1)}\left(2-\frac{t}{2}\right)- \left( \psi ^{(0)}\left(2-\frac{s}{2}\right)-\psi ^{(0)}\left(2-\frac{t}{2}\right) \right)^2 \right) \nonumber \\
& + R_1(s,t) \psi ^{(0)}\left(2-\frac{s}{2}\right)+R_1(t,s) \psi ^{(0)}\left(2-\frac{t}{2}\right)+ \pi^2 R_2(s,t)-\frac{32}{27 (s+t-8)} +a
\label{AdSbox}
\end{eqnarray}
where different choices of regularization would lead to different values of $a$ and the rational functions are given by
\begin{eqnarray}
R_0(s,t)&=& -\frac{4 \left(3 s^2 t-8 s^2+3 s t^2-32 s t+60 s-8 t^2+60 t-96\right)}{27 (s+t-8) (s+t-6) (s+t-4)}\;, \\
R_1(s,t)&=& \frac{8 \left(3 s^2+3 s t-26 s-10 t+48\right)}{27 (s+t-8) (s+t-4)}\;,\\
R_2(s,t) &=& \frac{4 \left(3 s^2 t-8 s^2+3 s t^2-32 s t+60 s-8 t^2+60 t-96\right)}{27 (s+t-8) (s+t-6) (s+t-4)}\;.
\end{eqnarray}
Given this explicit expression it is straightforward to compute the flat space limit where $-s$, $-t$ are large. We obtain
\begin{equation}
\mathcal{B}^{\rm 8d}_{st} \sim \frac{4 s t \log ^2\left(\frac{-s}{-t}\right)}{9 (s+t)^2} + \frac{8 (s \log (-s)+t \log (-t))}{9 (s+t)}+\frac{4 \pi ^2 s t}{9 (s+t)^2} + \beta\;,
\end{equation}
for some constant $\beta$ which depends on $a$. This precisely agrees with the eight dimensional box function in flat space! Note that also in this case the computation involves the introduction of a regulator, and we have a corresponding ambiguity. The method presented here can also be used to resum the $AdS_5\times S^5$ one-loop reduced Mellin amplitudes obtained in \cite{Alday:2018kkw,Alday:2019nin}.\footnote{Due to the different form of the $c_{mn}$ coefficients, this strategy does not immediately apply to the 5d amplitude (\ref{oneloopE8box}) with just the $AdS_5$ factor. However, we expect similar ambiguity structure which contains contact terms affecting only the CFT data of double-trace operators with low-lying spins.} 

\section{Coupling to gravity}\label{Sec:coupgravity}
In Section \ref{Sec:loopAdS5S3}, we considered the gluon one-loop contribution to the four-point function $\langle 2222\rangle$ which is of order $\mathcal{O}(N^{-2})$. However, at the same order there is another independent contribution arising from the tree-level exchange of the massless supergravity multiplet. Such contributions to super gluon four-point functions were considered for $AdS_6$ and $AdS_7$ theories in \cite{Zhou:2018ofp},\footnote{However, in these theories the super graviton exchange contributions are {\it not} of the same order as the super gluon one-loop contributions.} and the correlators were shown to be fixed by the superconformal Ward identities up to an overall normalization determined by the stress tensor central charge. Here we similarly compute the super graviton exchange amplitude for $AdS_5$.

The exchanged supergravity multiplet includes an $SU(2)_R$ singlet scalar super primary field of dimension 2, a vector field of dimension 3 and $SU(2)_R$ spin 1, and an $SU(2)_R$ neutral graviton field of dimension 4.\footnote{Note that supergravity multiplets of higher Kaluza-Klein levels are not exchanged in this correlator.} We can compute the amplitude due to super graviton multiplet exchange using the position space method developed in \cite{Rastelli:2016nze,Rastelli:2017udc}. We write down an ansatz for the four-point function  as the linear combination of all exchange diagrams and contact diagrams with no more than two derivatives
\begin{equation}
G_{2222,{\rm gravity}}^{I_1I_2I_3I_4}=\delta^{I_1I_2}\delta^{I_3I_4}G_s+\delta^{I_1I_4}\delta^{I_2I_3}G_t+\delta^{I_1I_3}\delta^{I_2I_4}G_u+G_{con}^{I_1I_2I_3I_4}\;.
\end{equation}
In this ansatz, $G_s$ is the s-channel exchange contribution
\begin{equation}
G_s=\frac{(v_1\cdot v_2)^2(v_3\cdot v_4)^2}{x_{12}^4x_{34}^4}\left(\lambda_s\mathcal{W}_{2,0}+\lambda_v (\alpha-\frac{1}{2})\mathcal{W}_{3,1}+\lambda_g \mathcal{W}_{4,2}\right)\;,
\end{equation}
where $(\alpha-\frac{1}{2})$ is an $SU(2)_R$ polynomial capturing the exchange contribution of the spin-1 contribution, and $\lambda_s$, $\lambda_v$, $\lambda_g$ are unknown parameters. The exchange Witten diagrams $\mathcal{W}_{\Delta,\ell}$ can be expressed as a finite sum of the $D$-functions, and formulae for computing such diagrams can be found in Appendix A of \cite{Rastelli:2017udc}. $G_t$ and $G_u$ are related to $G_s$ by crossing symmetry. We have assumed the contact part $G_{con}^{I_1I_2I_3I_4}$ contains all R-symmetry and color structures, {\it i.e.}, it is a degree-2 polynomial in $\alpha$ after extracting the factor $(v_1\cdot v_2)^2(v_3\cdot v_4)^2$ and contributes to all color channels. Furthermore, we assume that the contact term contains at most two derivatives, so that the correlator has the correct behavior in the flat space limit. This means we have only $D_{2,2,2,2}$, and $x_{12}^2D_{3,3,2,2}$ with its crossing images. Imposing the superconformal Ward identities (\ref{scfWardidposi}),  we can fix all coefficients in the ansatz up to an overall constant.\footnote{Equivalently, one could also write down  the ansatz in Mellin space, and then solve the ansatz by imposing the Mellin space superconformal Ward identities \cite{Zhou:2017zaw,Alday:2020dtb}.} The answer is given by 
\begin{equation}\label{Ggravity}
G_{2222,{\rm gravity}}^{I_1I_2I_3I_4}={\rm R}\, H_{2222,{\rm gravity}}^{I_1I_2I_3I_4}
\end{equation}
where 
\begin{equation}
H_{2222,{\rm gravity}}^{I_1I_2I_3I_4}=\frac{24(C_{2,2,g})^2}{\pi^2}\bigg(\delta^{I_1I_2}\delta^{I_3I_4}\frac{D_{2,2,3,3}}{x_{12}^2}+\delta^{I_1I_4}\delta^{I_2I_3}\frac{D_{2,3,3,2}}{x_{14}^2}+\delta^{I_1I_3}\delta^{I_2I_4}\frac{D_{2,3,2,3}}{x_{13}^2}\bigg)\;.
\end{equation}
The coefficient $C_{2,2,g}$ appears in the three-point function two $\mathcal{O}_2$ and the super primary of the massless supergravity multiplet as
\begin{equation}
\langle \mathcal{O}_2(x_1,v_1)\mathcal{O}_2(x_2,v_2)\mathcal{O}_g(x_3)\rangle=C_{2,2,g}\frac{(v_1\cdot v_2)^2}{x_{12}^2x_{13}^2x_{23}^2}\;,
\end{equation}
and is related to the stress tensor central charge by
\begin{equation}
(C_{2,2,g})^2=\frac{80}{3C_{\mathcal{T}}}\;.
\end{equation}
Here the central charge appears in the stress tensor two-point function as 
\begin{equation}
\langle\mathcal{T}_{\mu\nu}(x)\mathcal{T}_{\rho\sigma}(0)\rangle=\frac{C_{\mathcal{T}}}{4\pi^2 x^8}\left(I_{\mu\sigma}I_{\nu\rho}+I_{\mu\rho}I_{\nu\sigma}-\frac{1}{2}\delta_{\mu\nu}\delta_{\sigma\rho}\right)
\end{equation}
where $I_{\mu\nu}=\delta_{\mu\nu}-2\frac{x_\mu x_\nu}{x^2}$. In the theories of D3 probing F-theory singularities, the central charge is \cite{Aharony:2007dj}
\begin{equation}
C_{\mathcal{T}}=\frac{35N^2}{2\pi^2(2-\nu)}+\mathcal{O}(N)
\end{equation}
where $\nu$ characterizes the singularity type (see foonote \ref{fnsing}). 
Note an interesting feature of (\ref{Ggravity}) is that it does not contain a protected part. Under the twist $\alpha=1/z$ or $\alpha=1/\bar{z}$, the four-point correlator vanishes. 

Let us also express this correlator in Mellin space. The Mellin amplitude of the four-point function is
\begin{equation}
\mathcal{M}_{2222,{\rm gravity}}^{I_1I_2I_3I_4}=\delta^{I_1I_2}\delta^{I_3I_4}\mathcal{M}_s+\delta^{I_1I_4}\delta^{I_2I_3}\mathcal{M}_t+\delta^{I_1I_3}\delta^{I_2I_3}\mathcal{M}_u
\end{equation}
where 
\begin{equation}
\mathcal{M}_s=-3(C_{2,2,g})^2(v_1\cdot v_2)^2(v_3\cdot v_4)^2\bigg(\frac{(u-4)^2+2s(u-1)\alpha-8\alpha}{s-2}+(s-4)\alpha^2\bigg)\;,
\end{equation}
and $\mathcal{M}_t$, $\mathcal{M}_u$ can be obtained by using crossing symmetry. This Mellin amplitude can be further expressed in terms of a compact reduced Mellin amplitude defined from $H$
\begin{equation}
\widetilde{\mathcal{M}}_{2222,{\rm gravity}}^{I_1I_2I_3I_4}=-12(C_{2,2,g})^2\bigg(\frac{\delta^{I_1I_2}\delta^{I_3I_4}}{s-2}+\frac{\delta^{I_1I_4}\delta^{I_2I_3}}{t-2}+\frac{\delta^{I_1I_3}\delta^{I_2I_4}}{\tilde{u}-2}\bigg)
\end{equation} 
where we recall $\tilde{u}=u-2=6-s-t$.

\section{Discussion and outlook}\label{Sec:disc}
In this paper, we initiated the study of AdS gluon amplitudes at loop levels. Our main results are the simple Mellin space formulae (\ref{oneloopE8box}) and (\ref{oneloopAdS5S3}) for four-point one-loop amplitudes of SYM on $AdS_5$ and $AdS_5\times S^3$. For the case of $AdS_5\times S^3$, we performed the sums in the amplitude and obtained a closed form expression, see (\ref{AdSbox}). These results are reminiscent of the flat space amplitudes, with logarithms replaced by polygamma functions, and reproduce the latter in the large AdS radius limit. Our work leads to a number of natural research avenues for the future. We list some of them below.
\begin{itemize}
\item In this paper, we focused on the simplest one-loop amplitude with the lowest Kaluza-Klein level $k_i=2$. To further explore the loop-level dynamics, we should also study more general amplitudes with higher Kaluza-Klein levels. Computing these amplitudes requires a thorough analysis of the mixing problem at the tree level which is so far missing in the literature.
\item Relatedly, the tree-level super gluon amplitudes on $AdS_5\times S^3$ was shown to display a hidden eight dimensional conformal symmetry \cite{Alday:2021odx}. It would be interesting to explore how this structure can help to organize the loop-level amplitudes.
\item Recently, a double-copy-like relation was found in \cite{Zhou:2021gnu} which relates all tree-level four-point amplitudes of $AdS_5\times S^5$ IIB supergravity, $AdS_5\times S^3$ SYM, and bi-adjoint scalars on $AdS_5\times S^1$. It would be very interesting to see if the tree-level relation can be extended to the loop level as well. 
\item For super gluon four-point functions at $\mathcal{O}(N^{-2})$ the only loop contribution is the one which we computed in this paper with gluons running in the loop. However, at $\mathcal{O}(N^{-4})$ we can also have one-loop box diagrams where two internal legs are gluons and the other two legs are gravitons. Since the gluons are restricted to the eight dimensional subspace while the gravitons can propagate in the full ten dimensional space, it will be particularly interesting to examine the flat space limit of this amplitude. Presumably, it should match a ten dimensional flat space one-loop box integral where the two of the four internal propagators are confined to a codimension-2 subspace.  
\item It would also be interesting to extend the one-loop analysis to other backgrounds which have AdS factors other than $AdS_5$, and to explore the structure of one-loop amplitudes across different spacetime dimensions. Similar supergravity one-loop four-point amplitudes have been computed for eleven dimensional supergravity on $AdS_7\times S^4$ \cite{Alday:2020tgi} and $AdS_4\times S^7$ \cite{Alday:2021ymb}.
\item We can also apply the AdS unitarity method to compute all-loop contributions that correspond to the iterated s-cuts in flat space, as has been done in the supergravity case \cite{Bissi:2020wtv,Bissi:2020woe}. However, to fully determine these higher-loop correlators knowing just tree-level four-point functions is no longer sufficient. We would also need the information of multi-trace operators which are encoded in higher-point correlators. 
\end{itemize}

\acknowledgments
The work of L.F.A. is supported by funding from the European Research Council (ERC) under the European Union's Horizon 2020 research and innovation programme (grant agreement No 787185), and in part by the STFC grant ST/T000864/1. The work of A.B is supported by Knut and Alice Wallenberg Foundation under grant KAW 2016.0129 and by VR grant 2018-04438. The work of X.Z. in Princeton was supported in part by Simons Foundation Grant No. 488653. The work of X.Z. in China is supported by funds from University of Chinese Academy of Sciences (UCAS) and from the Kavli Institute for Theoretical Sciences (KITS).

\appendix

\section{$\mathcal{N}=2$ superconformal blocks}\label{scfblocks}
For reader's convenience, we record below the expressions for  superconformal blocks which were obtained in \cite{Nirschl:2004pa}.

\vspace{0.3cm}

\noindent{\it Long multiplet $\mathcal{A}^{\Delta}_{R,(\frac{\ell}{2},\frac{\ell}{2})}$:}
\begin{equation}
S^{\mathcal{A}}_{\Delta,R,\ell}=(y-x)(y-\bar{x})P_R(y)G_{\Delta+2,\ell}
\end{equation}
where $P_R(y)$ is the Legendre polynomial. 

\vspace{0.3cm}
\noindent{\it Semi-short multiplet $\hat{\mathcal{C}}_{R,(\frac{\ell}{2},\frac{\ell}{2})}$:}
\begin{eqnarray}
\nonumber S^{\mathcal{C}}_{R,\ell}=&&P_R(y)\big(G_{2R+\ell+2,\ell}+\tfrac{1}{4}a_{R}G_{2R+\ell+4,\ell}+a_{R+\ell+2}G_{2R+\ell+4,\ell+2}\big)\\
\nonumber &&+P_{R-1}(y)\frac{R}{2R+1}\big(G_{2R+\ell+3,\ell+1}+\tfrac{1}{4}G_{2R+\ell+3,\ell-1}+\tfrac{1}{4}a_{R+\ell+2}G_{2R+\ell+5,\ell+1}\big)\\
 &&+P_{R-2}(y)\frac{(R-1)R}{4(2R-1)(2R+1)}G_{2R+\ell+4,\ell}+P_{R+1}(y)\frac{R+1}{2R+1}G_{2R+\ell+3,\ell+1}
\end{eqnarray}
where 
\begin{equation}
a_R=\frac{R^2}{(2R-1)(2R+1)}\;.
\end{equation}

\vspace{0.3cm}
\noindent{\it Short multiplet $\hat{\mathcal{B}}_{R}$:}
\begin{equation}
S^{\mathcal{B}}_{R}=P_R(y)G_{2R,0}+\frac{R}{2R+1}P_{R-1}(y)G_{2R+1,1}+\frac{(R-1)R}{4(2R-1)(2R+1)}P_{R-2}(y)G_{2R+2,0}\;.
\end{equation}

Using these expressions, it is straightforward to extract the single variable function $f$ and the two variable function $\mathcal{K}$ in the representation (\ref{solscfWIfK}). In the singlet channel, we find (\ref{longK}), (\ref{fKC}) and (\ref{fKB}).

\section{Properties of $D$-functions}\label{App:Dfunctions}
In this appendix, we collect a few useful formulae of the $D$-functions. The $D$-functions are defined as a four-point contact Witten diagram in $AdS_{d+1}$ with no derivatives
\begin{equation}\label{defDf}
D_{\Delta_1,\Delta_2,\Delta_3,\Delta_4}(x_i)=\int \frac{d^dzdz_0}{z_0^{d+1}}\prod_{i=1}^4 G^{\Delta_i}_{B\partial}(z,x_i)\;,\quad G^{\Delta_i}_{B\partial}(z,x_i)=\left(\frac{z_0}{z_0^2+(\vec{z}-\vec{x}_i)^2}\right)^{\Delta_i}
\end{equation}
 Contact diagrams with derivatives can also be expressed as $D$-functions with shifted weights by using the identity
\begin{equation}
\nabla^\mu G^{\Delta_1}_{B\partial} \nabla_\mu G^{\Delta_2}_{B\partial}=\Delta_1\Delta_2(G^{\Delta_1}_{B\partial}G^{\Delta_2}_{B\partial}-2x_{12}^2G^{\Delta_1+1}_{B\partial}G^{\Delta_2+1}_{B\partial})\;.
\end{equation}
It is convenient to write the $D$-functions as functions of cross ratios by extracting a kinematic factor
\begin{equation}\label{dbar}
\frac{ \prod_{i=1}^4\Gamma(\Delta_i)}{\Gamma(\frac{1}{2}\Sigma_\Delta-\frac{1}{2}d)}\frac{2}{\pi^{\frac{d}{2}}}D_{\Delta_1,\Delta_2,\Delta_3,\Delta_4}(x_i)=\frac{(x_{14}^2)^{\frac{1}{2}\Sigma_\Delta-\Delta_1-\Delta_4}(x^2_{34})^{\frac{1}{2}\Sigma_\Delta-\Delta_3-\Delta_4}}{(x^2_{13})^{\frac{1}{2}\Sigma_\Delta-\Delta_4}(x^2_{24})^{\Delta_2}}\bar{D}_{\Delta_1,\Delta_2,\Delta_3,\Delta_4} (U,V)\, ,
\end{equation}
where $\Sigma_\Delta=\sum_{i=1}^4\Delta_i$. The $\bar{D}$-functions are independent of the spacetime dimension. 

The $D$-functions satisfy the derivative relations such as
\begin{equation}\label{DWS}
D_{\Delta_1, \Delta_2+1 \Delta_3+1, \Delta_4}(x_i)=\frac{d-\Sigma_\Delta}{2\Delta_2\Delta_3}\frac{\partial}{\partial x_{23}^2}D_{\Delta_1,\Delta_2, \Delta_3, \Delta_4}(x_i)\;.
\end{equation} 
We can rewrite the relations in terms of $
\bar{D}$-functions and get
\begin{equation}\label{DbarWS}
\begin{split}
\bar{D}_{\Delta_1+1,\Delta_2+1,\Delta_3,\Delta_4}={}&-\partial_U \bar{D}_{\Delta_1,\Delta_2,\Delta_3,\Delta_4}\;,\\
\bar{D}_{\Delta_1,\Delta_2,\Delta_3+1,\Delta_4+1}={}&(\Delta_3+\Delta_4-\tfrac{1}{2}\Sigma_\Delta-U\partial_U )\bar{D}_{\Delta_1,\Delta_2,\Delta_3,\Delta_4}\;,\\
\bar{D}_{\Delta_1,\Delta_2+1,\Delta_3+1,\Delta_4}={}&-\partial_V \bar{D}_{\Delta_1,\Delta_2,\Delta_3,\Delta_4}\;,\\
\bar{D}_{\Delta_1+1,\Delta_2,\Delta_3,\Delta_4+1}={}&(\Delta_1+\Delta_4-\tfrac{1}{2}\Sigma_\Delta-V\partial_V )\bar{D}_{\Delta_1,\Delta_2,\Delta_3,\Delta_4}\;,\\
\bar{D}_{\Delta_1,\Delta_2+1,\Delta_3,\Delta_4+1}={}&(\Delta_2+U\partial_U+V\partial_V )\bar{D}_{\Delta_1,\Delta_2,\Delta_3,\Delta_4}\;,\\
\bar{D}_{\Delta_1+1,\Delta_2,\Delta_3+1,\Delta_4}={}&(\tfrac{1}{2}\Sigma_\Delta-\Delta_4+U\partial_U+V\partial_V )\bar{D}_{\Delta_1,\Delta_2,\Delta_3,\Delta_4}\;.
\end{split}
\end{equation}
These identities allow us to shift the weights of $\bar{D}$-functions by taking derivatives. Another set of useful identities arise from the invariance of (\ref{defDf}) under permutation of operators. It is straightforward to find
\begin{equation}\label{Dbarpermutation}
\begin{split}
\bar{D}_{\Delta_1,\Delta_2,\Delta_3,\Delta_4} (U,V)=&V^{-\Delta_2}\bar{D}_{\Delta_1,\Delta_2,\Delta_4,\Delta_3} (U/V,1/V)\\
=&V^{\Delta_4-\frac{1}{2}\Sigma_\Delta}\bar{D}_{\Delta_2,\Delta_1,\Delta_3,\Delta_4} (U/V,1/V)\\
=&\bar{D}_{\Delta_3,\Delta_2,\Delta_1,\Delta_4} (V,U)\\
=&V^{\Delta_1+\Delta_4-\frac{1}{2}\Sigma_\Delta}\bar{D}_{\Delta_2,\Delta_1,\Delta_4,\Delta_3} (U,V)\\
=&U^{\Delta_3+\Delta_4-\frac{1}{2}\Sigma_\Delta}\bar{D}_{\Delta_4,\Delta_3,\Delta_2,\Delta_1} (U,V)\;.
\end{split}
\end{equation}

Let us now focus on the special $D$-function with $\Delta_i=1$. The associated $\bar{D}$-function is the well known scalar one-loop box integral in four dimensions, and evaluates to \cite{Usyukina:1992jd}
\begin{equation}\label{Phiinzzb}
\bar{D}_{1,1,1,1}\equiv\Phi(z,\bar{z})=\frac{1}{z-\bar{z}}\left(2{\rm Li}_2(z)-2{\rm Li}_2(\bar{z})+\log(z\bar{z})\log\big(\frac{1-z}{1-\bar{z}}\big)\right)\;.
\end{equation}
Using this explicit formula, it is straightforward to check that in the small $U$ expansion ({\it i.e.}, small $z$ and fixed $\bar{z}$), the $\log U$ coefficient of $\Phi(z,\bar{z})$ is given by 
\begin{equation}
\Phi(z,\bar{z})\big|_{\log U}=\frac{\log(1-z)-\log(1-\bar{z})}{z-\bar{z}}\;.
\end{equation}
Finally, from this expression we can verify the following differential recursion relations
\begin{equation}\label{Phidiffrecurapp}
\begin{split}
\partial_z\Phi(z,\bar{z})=&-\frac{\Phi(z,\bar{z})}{z-\bar{z}}+\frac{\log U}{(z-1)(z-\bar{z})}-\frac{\log V}{z(z-\bar{z})}\;,\\
\partial_{\bar{z}}\Phi(z,\bar{z})=&\frac{\Phi(z,\bar{z})}{z-\bar{z}}-\frac{\log U}{(\bar{z}-1)(z-\bar{z})}+\frac{\log V}{\bar{z}(z-\bar{z})}\;.
\end{split}
\end{equation}
These relations imply that any $\bar{D}$-function obtained from $\bar{D}_{1,1,1,1}$ by using the ``weight-shifting'' operators in (\ref{DbarWS}) can be written as a linear combination of the basis functions $\Phi(z,\bar{z})$, $\log U$, $\log V$ and $1$, with rational coefficients in $z$ and $\bar{z}$. This property was useful for manipulating the tree-level correlators in position space and extracting the protected part of correlators.

\section{Flat space limit of Mellin amplitudes}\label{App:flatspacebox}
Let us consider a holographic CFT in $1/c$ expansion. Its four-point correlators at one loop can be reconstructed following  \cite{Aharony:2016dwx}  and in a wide class of examples \cite{Alday:2018kkw,Alday:2019nin} they have a remarkable structure in Mellin space, namely, the Mellin amplitudes have only simultaneous simple poles. We consider a (partial)-Mellin amplitude of such a form
\begin{equation}
\label{Mellinpartial}
\mathcal{M}(s,t) = \sum_{m,n} \frac{c_{mn}}{(s-2m)(t-2n)}
\end{equation}
where $c_{mn}$ has the following large $m,n$ behavior
\begin{equation}
\label{largemn}
c_{mn} = \frac{(m n)^{\frac{D}{2}-3}}{(m+n)^{\frac{D}{2}-2}} + \cdots\;.
\end{equation}
We would like to obtain the flat space limit of $\mathcal{M}(s,t)$, defined as the limit where $s$, $t$ become large. Naively, we would expect $\mathcal{M}(s,t) 
\sim 1/(s t)$. However, this behavior is enhanced provided 
\begin{equation}
\sum_{m,n} \frac{(m n)^{\frac{D}{2}-3}}{(m+n)^{\frac{D}{2}-2}} = \text{divergent},
\end{equation}
which happens for $D > 4$. In this case, the leading contribution to the Mellin amplitude at large $s$, $t$ arises from the region with large $m,n \sim s,t$, and the Mellin amplitude in the flat space limit is well approximated by 
\begin{equation}
\mathcal{M}_{\rm flat}(s,t)=\int^\infty_0 dm dn \frac{(m n)^{\frac{D}{2}-3}}{(m+n)^{\frac{D}{2}-2}} \frac{1}{(s-2m)(t-2n)}\;.
\end{equation}
Note that this integral is real and finite for $s,t <0$ and $4<D<6$. We can compute $\mathcal{M}_{\rm flat}(s,t)$ in this region, and then analytically continue it to other regions. This integral can be performed as follows. First, we perform the integral over $m$ to obtain
\begin{equation}
\mathcal{M}_{\rm flat}(s,t)=- \int^\infty_0 dn \frac{2 n^{\frac{D}{2}-3} \, _2F_1\left(1,\frac{D-4}{2};\frac{D-2}{2};\frac{2 n}{t}+1\right)}{(D-4) t (2 n-s)}.
\end{equation}
The integral over $n$ can be performed by first using the integral representation for the hypergeometric function
\begin{equation}
\, _2F_1(a,b;c;z) = \frac{\Gamma (c) }{\Gamma (b) \Gamma (c-b)} \int_0^1 d\zeta \zeta ^{b-1} (1-\zeta  z)^{-a} (1-\zeta )^{c-b-1}
\end{equation}
and then integrating over $n$ and over $\zeta$, in that order. The result appears to be quite complicated, but using standard identities for hypergeometric functions it can be written in the following compact form
\begin{equation}
M_{\rm flat}(s,t)=\frac{\pi   \left(s^2 (-t)^{D/2} \, _2F_1\left(1,\frac{D}{2}-2;\frac{D}{2}-1;\frac{s+t}{s}\right)+t^2 (-s)^{D/2} \, _2F_1\left(1,\frac{D}{2}-2;\frac{D}{2}-1;\frac{s+t}{t}\right)\right)}{2^{\frac{D}{2}-3} (D-4) \sin \left(\frac{\pi  D}{2}\right) s^3 t^3}
\end{equation}
where the analytic continuation $s \to s+ i \epsilon,t \to t+i \epsilon$ is understood. We see that $M_{\rm flat}(s,t)$ is a function of $s/t$ up an overall factor
\begin{equation}
\mathcal{M}_{\rm flat}(s,t) \sim \frac{s^{D/2-3}}{t} f(s/t)
\end{equation}
 which indeed displays an enhanced behavior with respect to $1/(s t)$ for $D>4$. Furthermore, up to an overall coefficient, $\mathcal{M}_{\rm flat}(s,t) $ exactly agrees with the flat space massless scalar box integral in $D$ dimensions
\begin{equation}
\mathcal{M}_{\rm flat}(s,t)=\frac{2^{\frac{D}{2}-3} \Gamma \left(\frac{D-3}{2}\right)}{\sqrt{\pi }} I^{(D)}_{\rm box}(s,t)
\end{equation}
where
\begin{equation}
I^{(D)}_{\rm box}(s,t) = \Gamma \left(4-\frac{D}{2}\right) \int_0^1 \prod_{i=1}^4 da_i \delta(\sum_{i=1}^4 a_i-1)\left( -s a_1 a_3-t a_2 a_4 \right)^{\frac{D}{2}-4}\;. 
\end{equation}
In summary, the partial Mellin amplitude (\ref{Mellinpartial}) with the behavior (\ref{largemn}) reduces to the $D$-dimensional massless scalar box integral in the flat space limit.

\bibliography{refgluon} 
\bibliographystyle{utphys}
\end{document}